\documentclass[aps,prb,floatfix,twocolumn,showpacs]{revtex4}
\usepackage{latexsym}
\usepackage{graphicx}

\usepackage{subfigure} 
\usepackage{epsfig}
\usepackage{amsmath}
\begin{document}

\title{Hole    spectral   functions    in   lightly    doped   quantum
  antiferromagnets}   \author{Satyaki   Kar$^{(1)}$   and   Efstratios
  Manousakis$^{(1,2)}$  }  \affiliation{$^{1}$Department  of  Physics,
  Florida  State  University,  Tallahassee,  FL  32306-4350,  USA  and
  \\    $^{2}$Department   of    Physics,   University    of   Athens,
  Panepistimioupolis, Zografos, 157 84 Athens, Greece.}  
\date{\today}
\begin{abstract}
We study the hole and magnon  spectral functions as a function of hole
doping  in  the two-dimensional  (2D)  $t-J$  and $t-t'-t''-J$  models
working within the limits of  the spin-wave theory, by linearizing the
hole-spin-deviation  interaction  and  by  adapting  the  non-crossing
approximation (NCA).   We find that   the staggered magnetization
decreases rather rapidly with doping and it goes to zero at a few percent
of hole  concentration in  both $t-J$ and  the $t-t'-t''-J$  model. We
find  that with doping, the residue of the quasiparticle peak
at  $\vec G=(\pm\pi/2,\pm\pi/2)$ decreases rapidly with doping and
the spectral function is in agreement with high resolution 
angle-resolved photo-emission spectroscopy (ARPES) studies of  the copper-oxide
superconductors. The observed large shift of the chemical potential inside
the Mott gap is found to be a result of broadening of the quasiparticle
peak. We find pockets centered at $\vec G$,
 similar to those observed by quantum oscillation
measurements, (i)  with  an   elliptical  shape   with  large
eccentricity along the anti-nodal direction in the case of the $t-J$ model,
and  (ii) with an almost  circular shape in  the case  of  the $t-t'-t''-J$ 
 model. 
We also find that
the  spectral intensity distribution  in the doped  antiferromagnet has a 
waterfall-like patten  along the  nodal direction  of the  Brillouin  zone, a
feature that  is also seen  in ARPES measurements. 

\end{abstract}
\maketitle

\section{Introduction}

During the last two decades there is significant progress in angle-resolved 
photo-emission spectroscopy (ARPES) due to improved energy and angle resolution
by at least an order of magnitude\cite{damascellirmp,damascelli,graf,Inosov,waterfall1,waterfall2,zhou,shen,fournier}. Furthermore, advances in improving
sample quality and characterization have enabled ARPES to become a leading tool
in the investigation of otherwise hidden information about 
copper-oxide based superconducting (COSC)  materials. ARPES directly
exposes the momentum and frequency dependent imaginary part of 
the single particle Green's function, a quantity which sits at the heart
of quantum many-body theory. 

As the wave-number of the incident photon is varied, a  smooth  transition  
of the intensity resembling  a  waterfall-like pattern  is
observed in  ARPES 
 of  the COSC  materials    at   various    hole    or   electron
doping concentrations\cite{graf,Inosov,waterfall1,waterfall2,zhou}.  
While it has been argued
that this could be  a   matrix  element
effect\cite{susmita},  similar  intensity  features  are  obtained by means of
theoretical models and recently in different materials
and under various conditions\cite{fournier}. 
 Quantum Monte Carlo  calculations showed such   intensity    
pattern   in   2D   Hubbard models\cite{macridin}. In addition, 
it was suggested\cite{strings}  that these waterfall-like
features of the ARPES intensity are caused by the gradual transfer of 
the spectral weight from the lowest energy quasiparticle peak at the
wavevector
$\vec G=(\pm\pi/2,\pm\pi/2)$ to the higher energy so-called ``string'' states
as the incident wave-vector gradually moves towards $(0,0)$ along the
nodal direction of the Brillouin zone. This suggestion was based 
on a calculation of the single hole spectral function in the $t-J$ model 
by  {\it  broadening}  the  
peaks of the spectral function corresponding to the string 
excitations\cite{strings}.  Later, the broadening
required to achieve a reasonable agreement with the APRES spectra
was obtained by  considering  the  effects  of  optical
phonons at  finite temperatures\cite{satyaki}. 
In the present paper we use the $t-J$  and $t-t'-t''-J$  models 
within the limits of  the spin-wave theory by linearizing the
hole-spin-deviation  interaction  and  within  the  non-crossing
approximation (NCA) to study the effects of doping on the single-hole spectral
functions. We show that the {\it broadening} of the peaks corresponding 
to the high energy string excitations can be achieved naturally
as an effect of doping, thus, leading to
a smooth waterfall-like intensity distribution  along the nodal direction 
of the Brillouin zone. 

Sometime ago, 
ARPES studies\cite{wells} on the undoped parent COSC compounds
reveal  the existence of a relatively well-defined 
low-energy quasiparticle peak at wavevector ${\vec G}$ which follows a 
dispersion with a bandwidth of similar magnitude to that calculated
in Ref.~\onlinecite{liu}. Rather recently, however, Fournier et
al.\cite{fournier}, reported that the quasiparticle residue as 
a function of doping is essentially zero below approximately $10\%$ hole 
concentration. In addition, Shen et al.\cite{shen2} have reported 
ARPES measurements where the quasiparticle peak has a broad line
shape (much broader than the ARPES resolution)
and they also found  what they call a ``paradoxical shift'' 
of the chemical potential within the Mott gap.
 Our results indicate that these seemingly contradictory 
findings can be reconciled using the calculations reported in the 
present paper. We  find that the lowest energy quasiparticle peak observed near
$(\pm \pi/2, \pm \pi/2)$ in the undoped cuprates\cite{wells}, is very
sensitive to a small amount of doping. Namely, the quasiparticle residue 
rapidly goes to zero as a function of doping concentration, 
and, in addition, it is known that optical phonons broaden 
the peak\cite{satyaki}.  Furthermore,
we find that the facts, that the quasiparticle peak is broad 
and that the chemical potential is shifted relative to the quasiparticle
peak position, are not two independent effects. When 
we repeated our calculation by broadening the quasiparticle
peak we found that the broadening itself leads to a very significant 
shift of the chemical potential relative to the lowest energy quasiparticle
position which can be understood in a straightforward manner.

Furthermore, we find that, at low doping, 
elliptical hole pockets form, which are centered around 
$\vec G$,  in agreement with rather recent 
quantum oscillation
measurements\cite{doiron}. Our calculated intensity of the spectral
function shows that there are luminous elliptical rings centered
around the points $\vec G$ of the Brillouin zone. We argue that
these rings could be related to the banana-like features seen in 
ARPES reported on the under-doped COSC materials\cite{shen}. Our 
formulation, which is based on the existence of antiferromagnetic (AF)
long-range-order (LRO), cannot yield a spectrum sharing the 
single-banana-like feature 
seen in ARPES,  where only one side around the $\vec G$
is luminous. The reason lies in the reflection symmetry about 
the line $(\pi,0)\to (0,\pi)$ 
of the Brillouin zone which is present when AF-LRO is present in the system.
Therefore, we view the banana-like shape as a result produced by 
``scissoring''  the elliptical ring along the above symmetry line, an
operation which breaks this symmetry required by the AF-LRO. 
The measurements presented in Ref.~\onlinecite{shen} were carried out
in materials which lack AF-LRO.

Linear  spin-wave (LSW) theory  can describe  the undoped  and lightly
doped antiferromagnets as  long as  AF-LRO is present
in  the system. In the  undoped COSC parent compounds,  which are 
antiferromagnetic insulators\cite{RMP},  the AF-LRO  disappears at a
very low doping concentration. For example,  in
$La_{2-x}Sr_xCuO_4$,  the  AF-LRO   is    destroyed   at    doping   $x\sim$
0.015\cite{keimer}. On the other hand, magnetic fluctuations persist  
even at higher doping concentrations  and short-range  spin 
correlations  are observed  in the
inelastic  neutron  scattering experiments  at  reasonably high  doping
concentrations\cite{keimer,PRL681414,PRB553886}.    Calculations   on
$t-J$ and related models also  show the staggered magnetization of the
system to  go to zero  as the undoped  system is gradually  doped with
holes\cite{vojta,horsch,gan,orbach2,belkasri}.

In order to study the 2D hole-doped antiferromagnet, we have considered the
$t-J$  and   $t-t'-t''-J$  models  
working within the limits of  the spin-wave theory, by linearizing the
hole-spin-deviation  interaction  and  by  adapting  the NCA.  
Within the NCA
the self-energy diagrams are considered for both holes and magnons and
the  Dyson's equation  is solved  to obtain  the  one-particle Green's
function.  The contribution  of  the magnon  polarization bubbles  are
considered  and, thus,  the magnon-mediated hole-hole interaction  is  
included. We  should mention  here that  we do not  use the  rigid band
approximation which uses the hole bands of the single hole calculation
to obtain the  hole spectra at finite concentration  neither do we use
any approximate  form of the hole  spectral function to write  it as a
sum  of a  coherent quasi-particle  peak and  an  incoherent continuum
\cite{fulde,orbach1}.

Hole and magnon Green's functions for 
a 2D $t-J$ model at finite-doping, using LSW approximation and NCA, 
have been studied  
earlier (Ref. \onlinecite{sherman}). 
However, the goal and focus of the present study is quite different; 
as already discussed, there are rather recent high  resolution ARPES 
studies which exhibit the existence of the waterfall 
features\cite{graf,Inosov,waterfall1,waterfall2,zhou} and
other studies which reveal the shape of the most intense part of the 
Brillouin zone near the
Fermi surface\cite{shen,fournier}.  Our calculations are done with the
aim to understand, if possible, these and related features
of the ARPES measurements. In addition, as we will discuss
in the paper, our results differ
significantly from the results of Ref.~\onlinecite{sherman}. For example,
we find that only a few percent hole concentration 
is enough to destroy the long-range antiferromagnetic order, while
in Ref.~\onlinecite{sherman} results for the spectral function were
 presented for up to hole doping $x\sim 0.25$, where we find that this
approach breaks down precisely because  there is no LRO for 
this level of doping.  

Our approach discussed above relies on the existence of 
AF-LRO both within the model and in the materials which we plan
to apply our findings, therefore, we only
hope to capture the physics of the real materials in the limit of light doping.
It is known, however, that even in the above models the AF-LRO may be
unstable upon doping leading to other forms of order, such as 
spiral order\cite{shraiman,jayaprakash,fulde}, stripe 
ordering\cite{white,hellberg} 
or phase separation instability\cite{emery,hellberg2}. 
Therefore, we begin by assuming that the AF-LRO is present
in the system at low doping and we proceed with a perturbative approach which
asserts that spin wave excitations are well-defined in the
system. If the AF order ceases to exist and it gets replaced by 
any of the above phases, this should show up as an instability
in our calculation. We will discuss this further in Sec.~\ref{magnetic}.

The paper is organized as follows:
In the next section we discuss the formulation and
the approximation scheme followed in this work. In Sec.~\ref{details}
of the Appendix
we outline some of the technical details of the calculation. 
In Sec.~~\ref{Holes} we present our results on the hole spectral functions,
the hole energy momentum dispersions and the intensity of the 
spectral functions is compared with the waterfall and banana-like features
observed in ARPES. In Sec.~\ref{magnons} our results on the magnon spectral 
functions, the
magnon
dispersion relations, and the staggered magnetization as a function of doping
are presented. In Sec.~\ref{conclusions} we summarize our findings and 
conclusions.

\section{Formulation}

The  2D $t-J$ or the $t-t'-t''-J$ model, when expressed in terms of 
operators which create spin-deviations above the N\'eel order and by keeping
up to quadratic terms in these operators, can be approximated by the 
following Hamiltonian\cite{liu}
\begin{eqnarray}
H&=&E_{0}+\sum_{\bf   k}\epsilon_k  (f^{\dagger}_{\bf  k}   f_{\bf  k}
+h^{\dagger}_{\bf k}  h_{\bf k})+ \sum_{\mathbf{k}}\Omega_{\mathbf{k}}
(\alpha_{\mathbf{k}}^{\dagger}\alpha_{\mathbf{k}}+
\beta_{\mathbf{k}}^{\dagger}\beta_{\mathbf{k}})\nonumber
\\       &+&\sum_{\mathbf{k},\mathbf{q}}h_{\mathbf{k}}^{\dagger}f_{\mathbf{k-q}}
[g(\mathbf{k},\mathbf{q})\alpha_{\mathbf{q}}+g(\mathbf{k-q},-\mathbf{q})
  \beta_{-\mathbf{q}}^{\dagger}]\nonumber\\ &+&f_{\mathbf{k}}^{\dagger}h_{\mathbf{k}
-\mathbf{q}}[g(\mathbf{k-q},-\mathbf{q})\alpha_{-\mathbf{q}}^{\dagger}+
g(\mathbf{k},\mathbf{q})\beta_{\mathbf{q}}]+H.c.
\label{linearized}
\end{eqnarray}
where  $h_{\mathbf{k}},f_{\mathbf{k}}$ are  the  spin-less hole  
annihilation operators  (the
so-called slave fermions) for $\downarrow$ and $\uparrow$ sub-lattices
respectively  and  $\alpha_{\mathbf{k}},\beta_{\mathbf{k}}$  
 are  the  magnon  annihilation
operators  for the  corresponding  sub-lattices. $\epsilon_k,~\Omega_k$  are
bare  hole and  magnon energies  respectively with  $\epsilon_k=0$ for
$t-J$            model            and            $\epsilon_k=4t^\prime
$cos$~k_x$cos$~k_y+2t^{\prime\prime}[$cos$~2k_x$+cos$~2k_y]$        for
$t-t^\prime-t^{\prime\prime}-J$   model.   We   use   the   parameters
$t^\prime=-0.33t$  and $t^{\prime\prime}=0.22t$  that  gives good  fit
with  ARPES results  of the  energy band  along anti-nodal 
directions \cite{Ronning}. The coupling  constants $g({\bf k},{\bf q})$ 
are defined in Ref. \onlinecite{liu}.
\begin{figure}[htp]
\vskip .2 in
\begin{center}
\includegraphics[width=.7\linewidth]{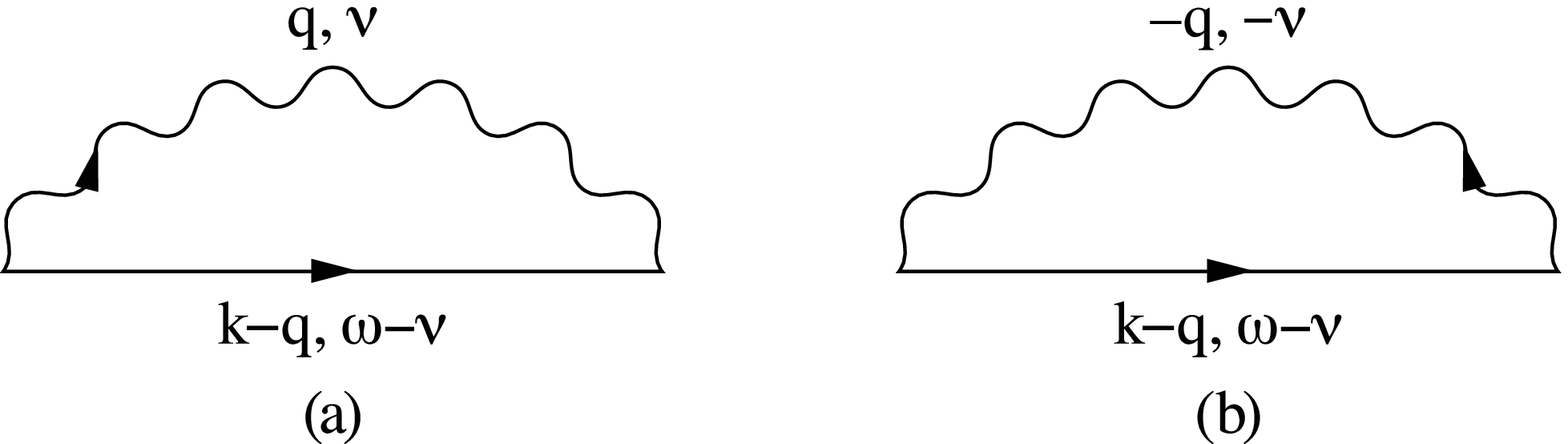}\\ \includegraphics[width=.3\linewidth]{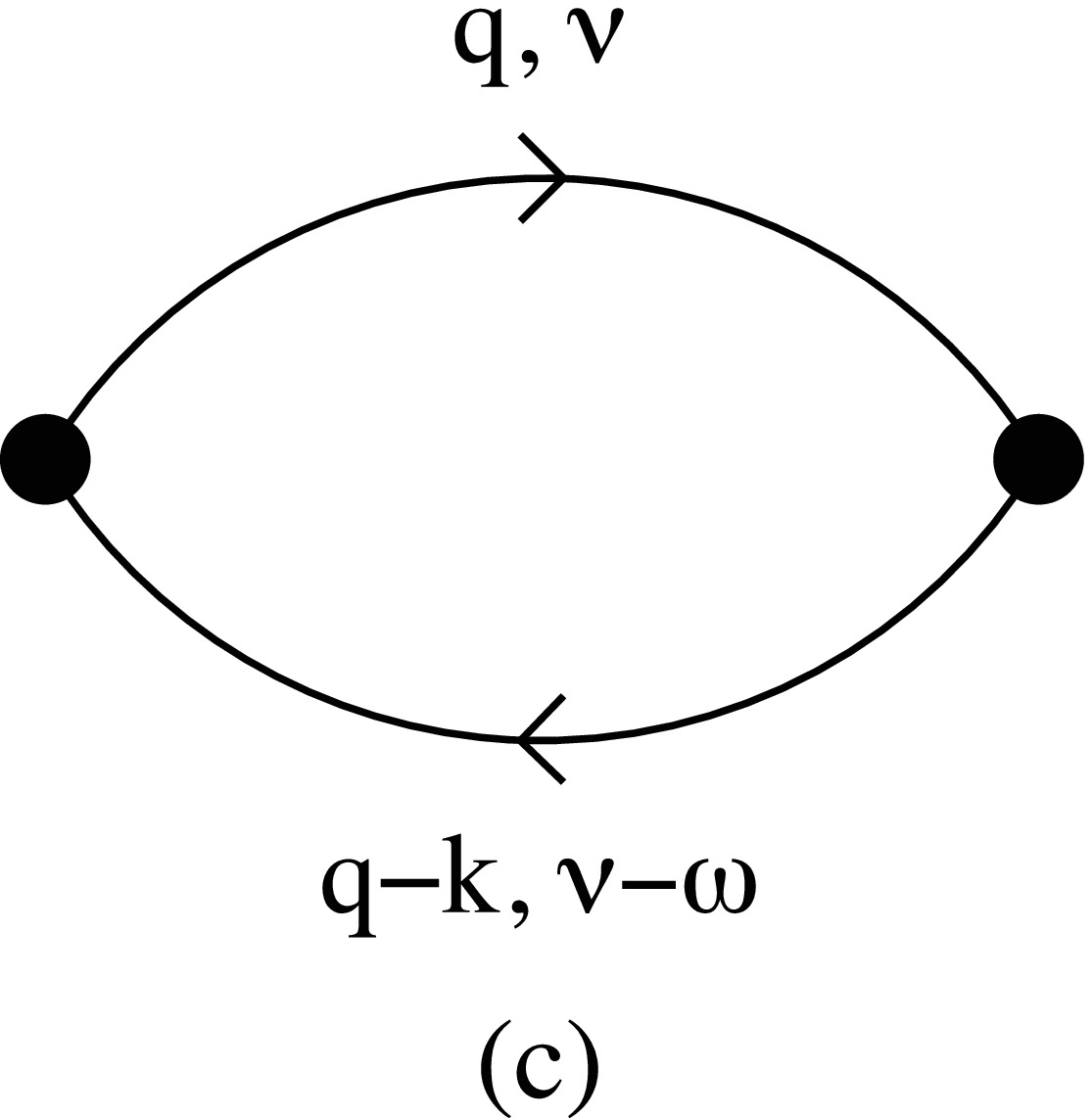}
\end{center}
\caption{NCA self-energy diagrams for holes (a,b) and magnons (c).}
\label{selfd}
\end{figure}

For   a  system   doped  with   holes,  the   Dyson's   equations  for
hole/spin-wave excitations become
\begin{eqnarray}
&&{G}({\bf k},\omega)=\frac{1}{\omega-\epsilon_{ {\bf k}}-
\Sigma({\bf k},\omega)+i~
    sgn(\omega-\mu)\eta}\nonumber,\\&&D({\bf k},\omega)=
\frac{1}{\omega-\Omega_{\bf k}-\Pi({\bf k},\omega)+i\eta},
\label{eqd1}
\end{eqnarray}
where $G({\bf k},\omega)$ and $D({\bf k},\omega)$  are the hole and 
magnon Green's
functions  and  $\Sigma({\bf k},\omega)$  and 
$\Pi({\bf k},\omega)$
are the  respective self-energies.  $\mu$ is the  chemical potential
for holes and $\eta$ is the broadening parameter. Here we would like
to note  that for a doping concentration  of $x$ in a  system of total
number  of  sites $N$,  the single-hole  Green's  function describes  the
spectral properties of a hole (or rather a
fermionic  excitation)  added  or  removed  from  the system  where  $xN$
number of holes  are already present.

In this  paper we  will work within the so-called 
non-crossing approximation (NCA), as the higher order vertex corrections
have been shown earlier to be small \cite{liu,liu2}.

The NCA self-energies are given by
\begin{eqnarray}
\Sigma({\bf k},\omega)&=&\frac{i}{2\pi}\sum_{\bf q}\int                        d\nu
      [g^2({\bf k},{\bf q})G({\bf k-q},\omega-\nu)D({\bf q},\nu)
\nonumber\\&&+g^2({\bf k-q},{\bf -q})G({\bf k-q},\omega-\nu)
D(-{\bf q},-\nu)],\nonumber
      \\     \Pi({\bf k},\omega)&=&-\frac{i}{2\pi}\sum_{ {\bf q}} 
g^2({\bf q}, {\bf k})\int    d\nu
      G({\bf q-k} ,\nu-\omega)G({\bf q},\nu).\nonumber\\
\label{eqd2}
\end{eqnarray}
The two terms of the hole self-energy describe the hole motion outside
the  Fermi-sea (of  holes) and  electron motion  inside  the Fermi-sea
respectively. Fig \ref{selfd}(a) describes hole motion forward in time
where  as Fig \ref{selfd}(b)  describes hole  motion backward  in time
which  is equivalent to  electron motion  forward in  time. It  can be
easily shown that this second term becomes zero for an undoped system.

Fig \ref{selfd}(c)  shows the magnon self-energy or spin-polarization
bubble. This  term is  also non-zero only  for the system  with finite
doping.

We can get rid of the infinite limits of the energy integral over $\nu$ if
we utilize the spectral representations of the Green's functions:
\begin{eqnarray}
&&\mathcal{G}({\bf k},\omega)=\int_{-\infty}^\infty
  d\epsilon[\frac{A({\bf k},\epsilon)}{\omega-\epsilon-\mu+i\eta}+
\frac{B({\bf k},\epsilon)}{\omega-\epsilon-\mu-i\eta}],\\
&&D({\bf k},\omega)=\int_{-\infty}^\infty
  \frac{C({\bf k},\epsilon)d\epsilon}{\omega-\epsilon+i\eta},
\label{dkw}
\end{eqnarray}
where $\mu$  is Fermi
energy and $A({\bf k},\epsilon)$, $B({\bf k},\epsilon)$  are hole  
spectral functions defined as
\begin{eqnarray}
&&A(k,\omega)=\sum_n\mid\langle n|h_k^\dagger|\psi\rangle\mid^2 \delta(\omega-\epsilon_n^{N+1}),~~~~~\nonumber\\
&&B(k,\omega)=\sum_m\mid\langle m|h_k|\psi\rangle\mid^2 \delta(\omega+\epsilon_m^{N-1}),
\label{eqb4}
\end{eqnarray}
where  $\{|n\rangle\}$ and $\{|m\rangle\}$ denote the complete set of 
basis states for ($N+1$) and ($N-1$) hole/particle systems
and $\epsilon_n^{N+1}$ and $\epsilon_m^{N-1}$ are the corresponding energies.
Here $|\psi\rangle$ is the interacting ground state of the
Hamiltonian. $C({\bf k},\epsilon)$  is the magnon  spectral function defined
as
\begin{eqnarray}
&&C(k,\omega)=\sum_n\mid\langle n|\alpha_k^\dagger|\psi\rangle\mid^2 \delta(\omega-\omega_n),
\label{eqb9}
\end{eqnarray}
where $\omega_n$ is the energy of the state 
$|n \rangle$ which can be connected to the ground state via the magnon
creation operator.

Using these auxiliary functions,  
the imaginary part of the  self-energy takes the following form
\begin{widetext}
\begin{eqnarray}
\Sigma_I({\bf k},\omega)&=&-\frac{1}{\pi}\sum_{{\bf q}}[\int_{0}^{\omega-\mu}    d\nu
  g^2({\bf k},{\bf q})
  G_I({\bf k-q},\omega-\nu)D_I({\bf q},\nu)\theta(\omega-\nu-\mu)+
\int^{0}_{\omega-\mu}d\nu
  g^2({\bf k-q},-{\bf q})\nonumber\\&&
  G_I({\bf k-q},\omega-\nu)D_I(-{\bf q},-\nu)\theta(\mu-\omega+\nu)]~,~~~~\Pi_I(k,\omega)=\frac{1}{\pi}[\int_{\mu-\omega}^\mu
  d\nu\sum_qg^2({\bf q},{\bf k})
  G_I({\bf q} ,\omega+\nu)G_I({\bf q-k},\nu)\nonumber\\ 
&&\theta(\omega)+\int_\mu^{\mu-\omega}
  d\nu\sum_qg^2({\bf q},{\bf k}) G_I({\bf q} ,\omega+\nu)G_I({\bf q-k},\nu)\theta(-\omega)].
\label{eqd5}
\end{eqnarray}
\end{widetext}

The derivation  of Eq. \ref{eqd5}  involves a convolution  of spectral
functions     utilizing    the   following representation for
the  delta     function: 
$\delta(x)=\frac{1}{\pi}lim_{\eta\to 0}\frac{\eta}{x^2+\eta^2}$.

The real parts are obtained by the Kramers-Kronig relation
\begin{eqnarray}
\begin{pmatrix}\Sigma_R({\bf k},\omega)\\\Pi_R({\bf k},\omega)
\end{pmatrix}=\frac{1}{\pi}P\int_{-\infty}^\infty\frac{ d\omega^\prime}{\omega^\prime-\omega}\begin{pmatrix}sgn(\omega^\prime-\mu)\Sigma_I({\bf k},\omega^\prime)
\\sgn(\omega^\prime)\Pi_I({\bf k},\omega^\prime)\end{pmatrix}.\nonumber\\
\label{eqd6}
\end{eqnarray}

Here, we should  mention that though the calculation  of the imaginary part
of  the self-energy avoids  the infinite  energy  integration limits,
the calculation of  the real part by Kramers-Kronig  relation does involve
an  energy  integration  between  $(-\infty,\infty)$  (which  is  done
numerically in the  range $(-8t,8t)$ as discussed in Sec.~\ref{details}).  
The sum
over $\vec q$ is avoided in the calculation of the real part.

Once  the   Green's  functions  are   obtained  we  can   compute  the
sub-lattice/staggered  magnetization which  is a  measure of  the long
range antiferromagnetic magnetic  order in the system. It  is given by
$M_S=S-\epsilon$   where  $S$   is  the   spin  of   the   system  and
$\epsilon=\frac{1}{N}\sum_k\left(\langle a_{\mathbf{k}}^{\dagger}
a_{\mathbf{k}}\rangle+\langle b_{\mathbf{k}}^{\dagger} b_{\mathbf{k}}\rangle\right)$,
the spin deviation and $a^{\dagger}_{\bf k}$ and 
$b^{\dagger}_{\bf k}$ the spin-deviation operators. 
For an undoped isotropic antiferromagnet this leads to
\begin{eqnarray}
\epsilon&=&\frac{1}{N}\sum_{\bf k}
\left(\frac{1}{\sqrt{1-\gamma_{\bf k}^2}}-1\right),
\label{eqd7}
\end{eqnarray}
where $\gamma_{\bf k}=(cos k_x+cos k_y)/2$.  For a 2D spin-1/2 isotropic
antiferromagnet  its value is  $\epsilon\sim0.197$. We  should mention
here that the validity of the  linear spin wave theory is based on the
smallness of the parameter $\epsilon$ \cite{RMP}.

In the case of the doped system  we also have 
$M_S=S-\epsilon$ where we find that
\begin{eqnarray}
\epsilon&=&\frac{1}{N}\sum_{\bf k} 
[\eta_{\bf k} (\langle\alpha_k\alpha_k^\dagger\rangle+\langle\beta_{-k}\beta_{-k}^\dagger\rangle) 
-(\eta_{\bf k} +1) \nonumber\\ &-&
\gamma_{\bf k}\eta_{\bf k}(\langle\alpha_k^\dagger\beta_{-k}^\dagger\rangle+\langle\beta_{-k}\alpha_{k}\rangle)],\\
\eta_{\bf k} &=& \frac{1}{\sqrt{1-\gamma_k^2}}.
\end{eqnarray}

Now, the relationship
\begin{eqnarray}
\langle\alpha_k\alpha_k^\dagger\rangle&=&\langle\beta_{-k}\beta_{-k}^\dagger\rangle=i~D(k,t\rightarrow0^{+})\\
\end{eqnarray}
leads to  
\begin{eqnarray}
\langle\alpha_k\alpha_k^\dagger\rangle&=&\int_{-\infty}^\infty
C(k,\epsilon)d\epsilon,
\end{eqnarray}
i.e., this is the total spectral weight of the magnon spectral function.

We  are  not   considering  the  off-diagonal  magnons,  i.e.,
$\langle\beta_{-k}\alpha_{k}\rangle=\langle\alpha_k^\dagger\beta_{-k}^\dagger\rangle=0$). Thus, the final expression is 
\begin{eqnarray}
\epsilon=\frac{1}{N}\sum_k\Big[\frac{1}
{\sqrt{1-\gamma_k^2}}\Big(
2\int C(k,\omega)d\omega-1\Big)-1\Big].
\end{eqnarray}

The technical part of the calculation is discussed in Sec.~\ref{details} of
appendix.

\section{Hole spectral functions}
\label{Holes}
\subsection{Spectral functions}
\label{spectral-functions}

\begin{figure}[htp]
\begin{center}
\vskip .3 in
\begin{tabular}{cc}
\epsfig{file=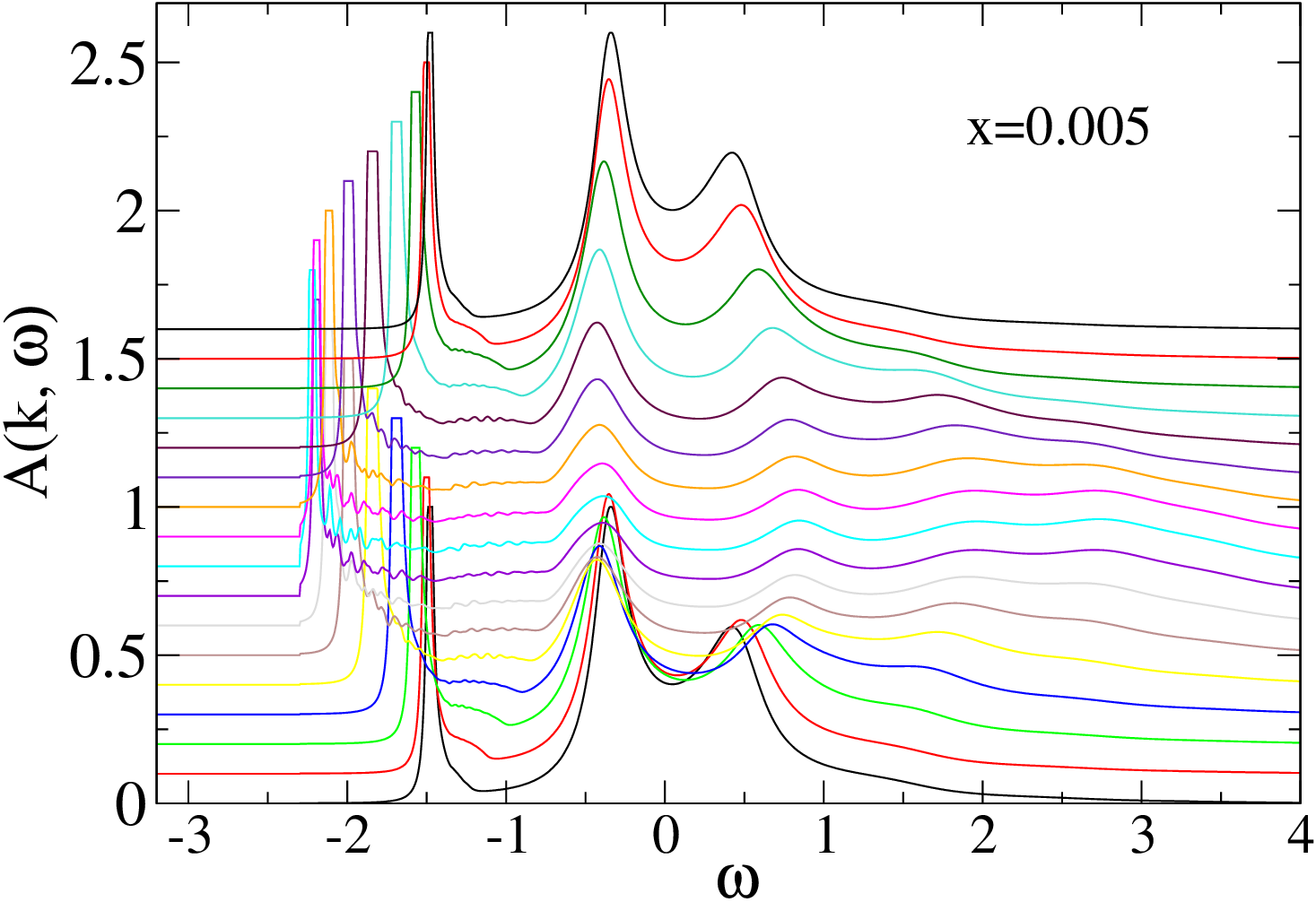,width=0.5\linewidth,clip=}&
\epsfig{file=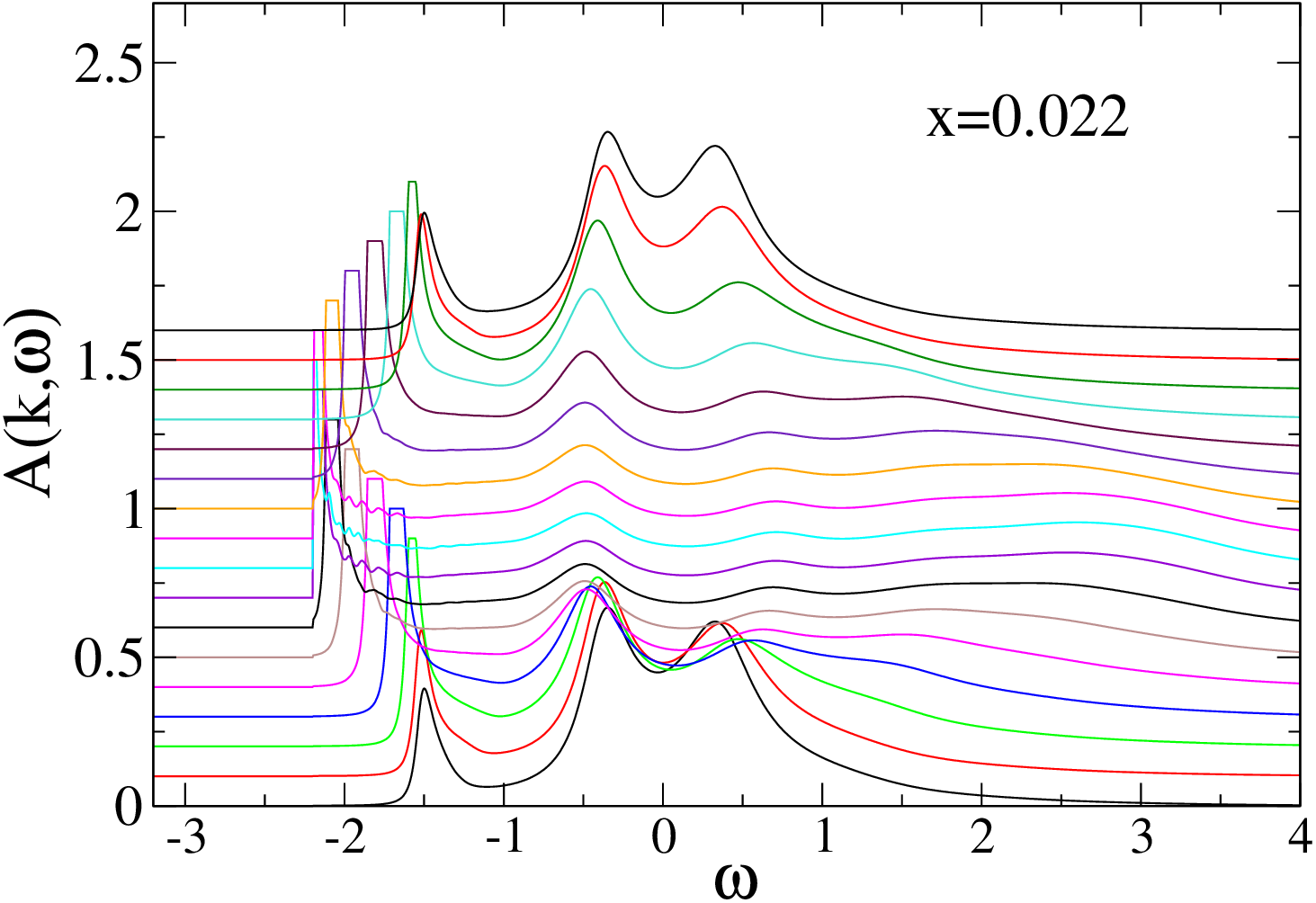,width=0.5\linewidth,clip=}\\ 
\epsfig{file=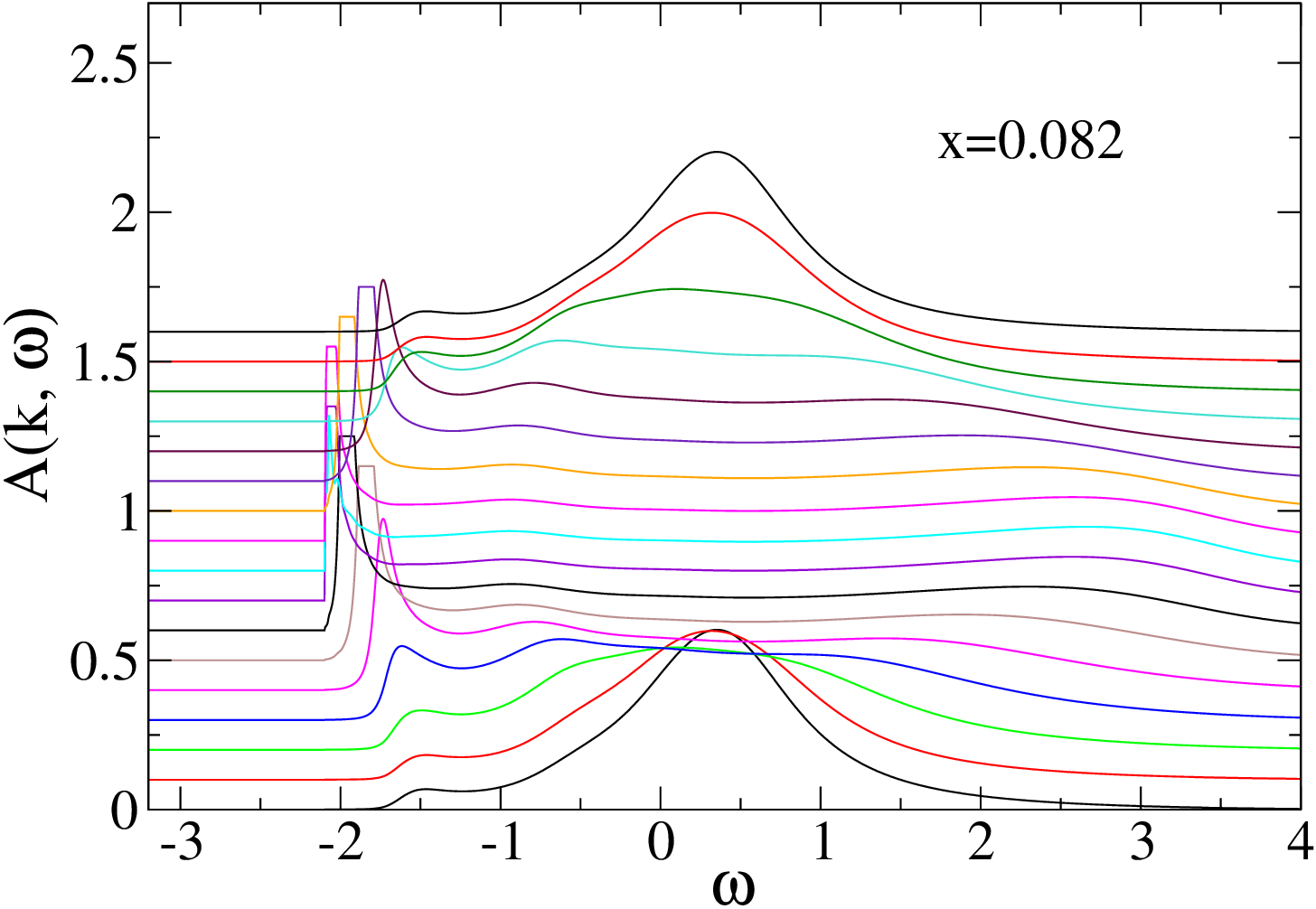,width=0.5\linewidth,clip=}&
\epsfig{file=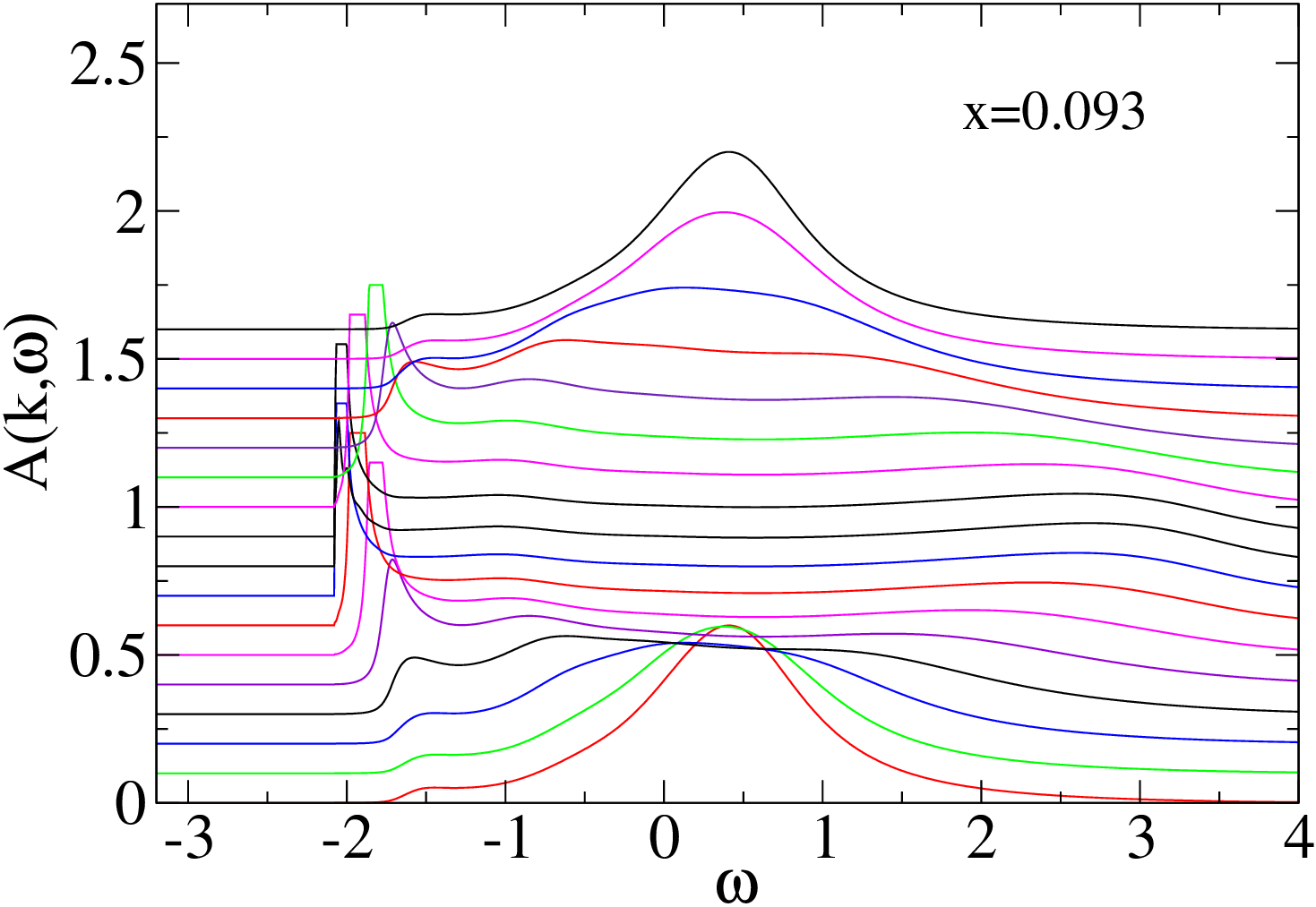,width=0.5\linewidth,clip=}
\end{tabular}
\end{center}
\caption{Hole  spectral  function of the 2D $t-J$ model 
at $J=0.4t (L=32)$ for various doping concentrations $x$.
The wavevector is
varied along the diagonal $(0,0) \to (\pi,\pi)$ direction of the Brillouin
zone. }
\label{figd2}
\end{figure}

\begin{figure}[htp]
\begin{center}
\vskip .3 in
\begin{tabular}{cc}
\epsfig{file=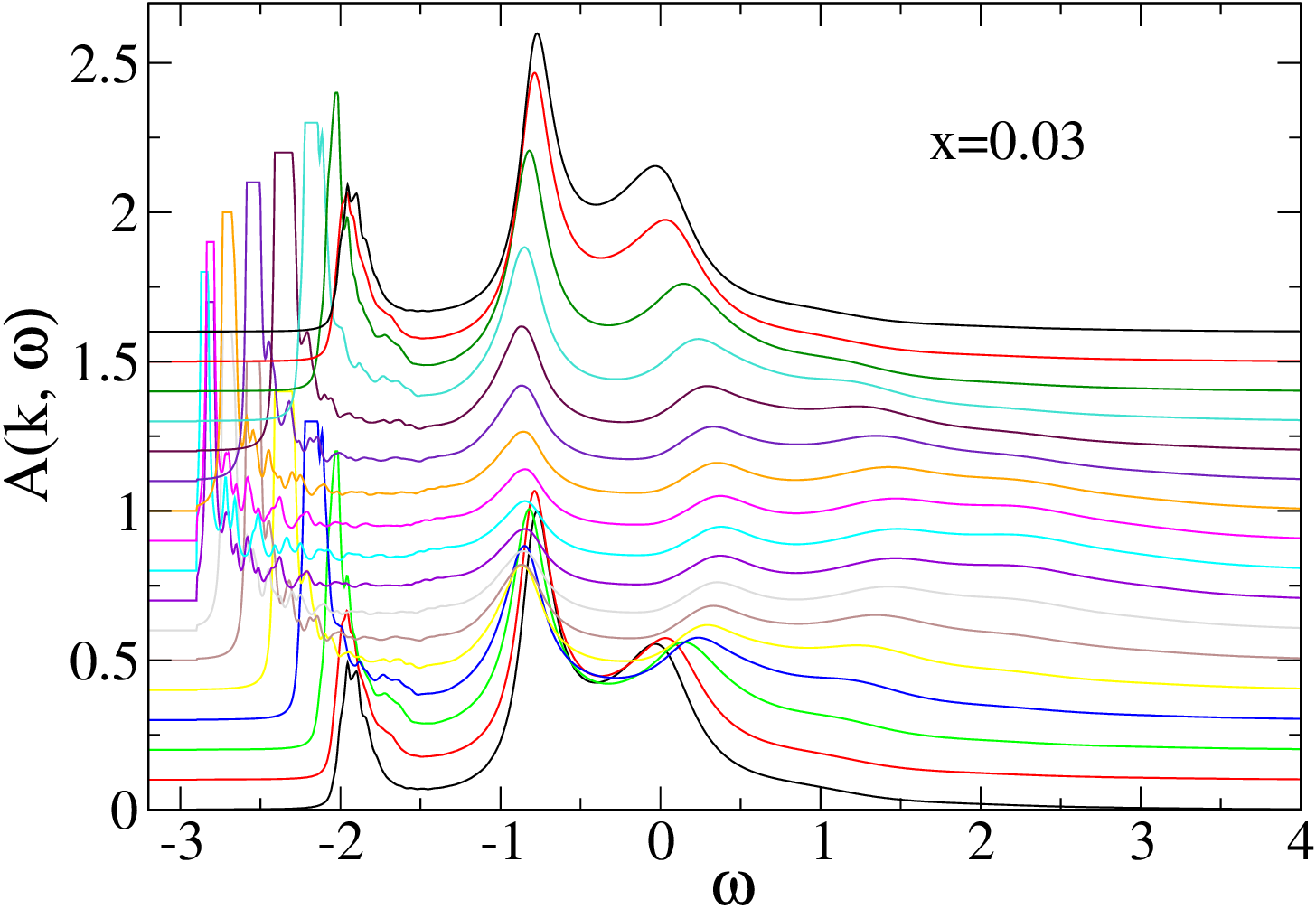,width=0.5\linewidth,clip=}&
\epsfig{file=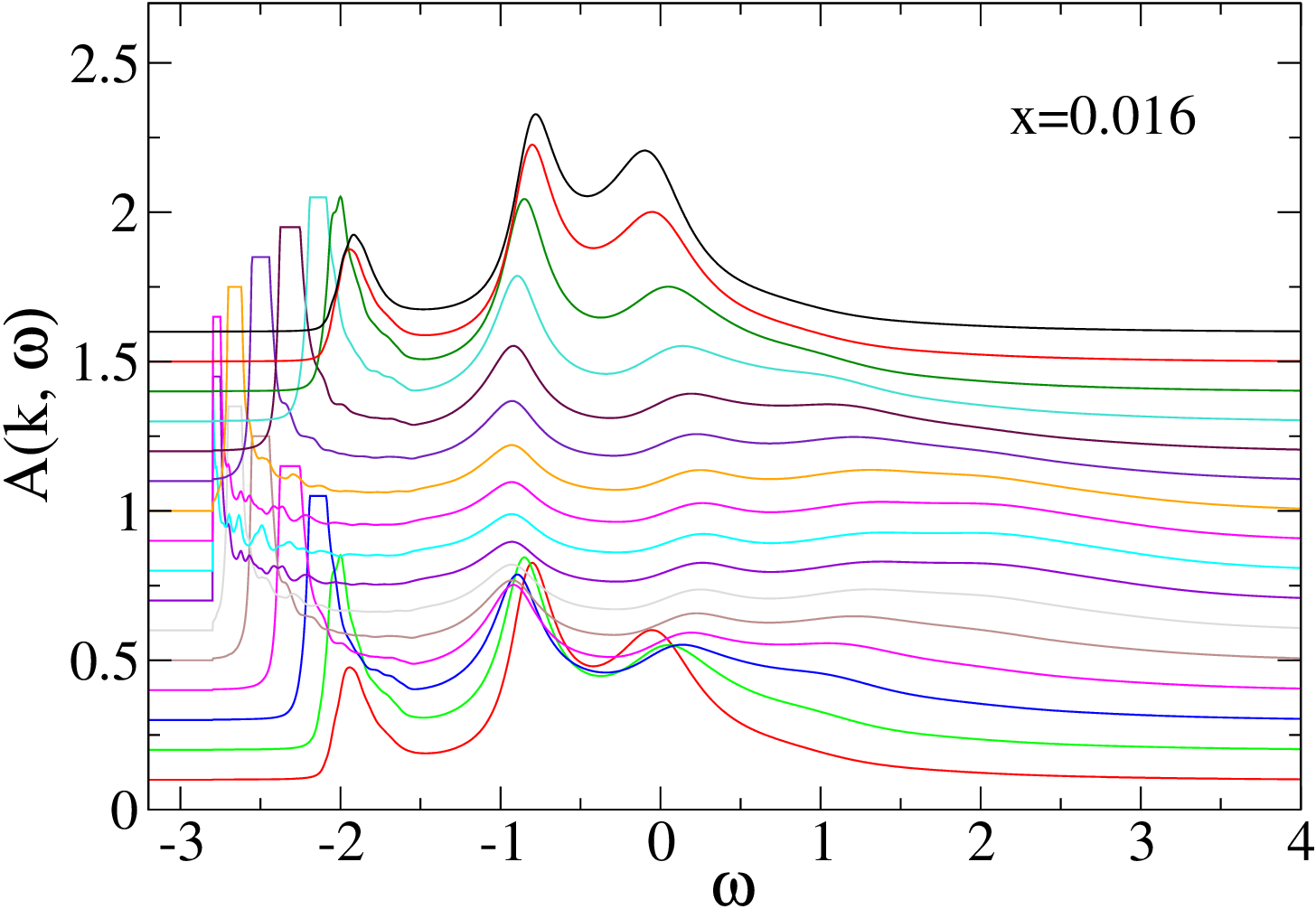,width=0.5\linewidth,clip=}\\ 
\epsfig{file=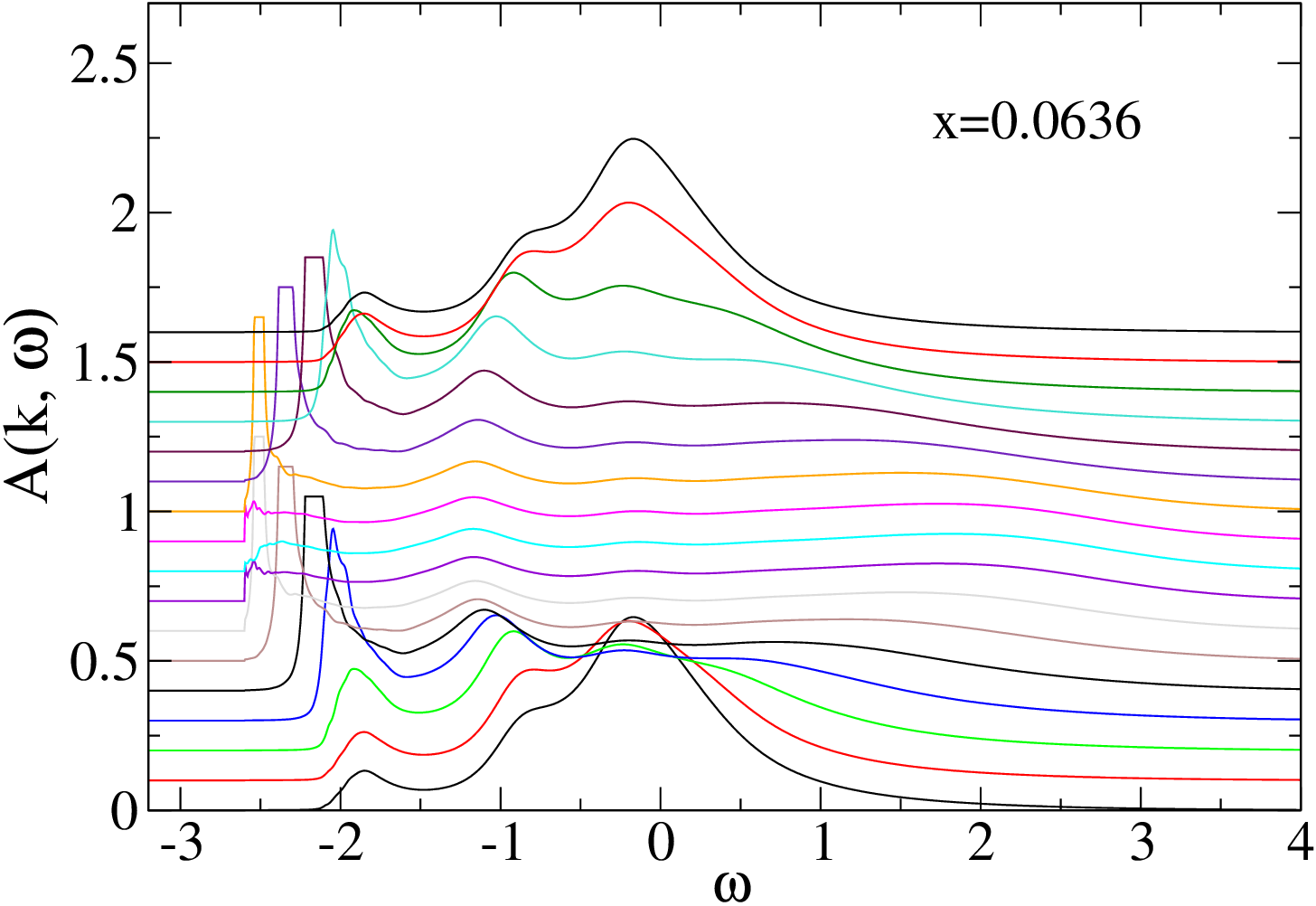,width=0.5\linewidth,clip=}&
\epsfig{file=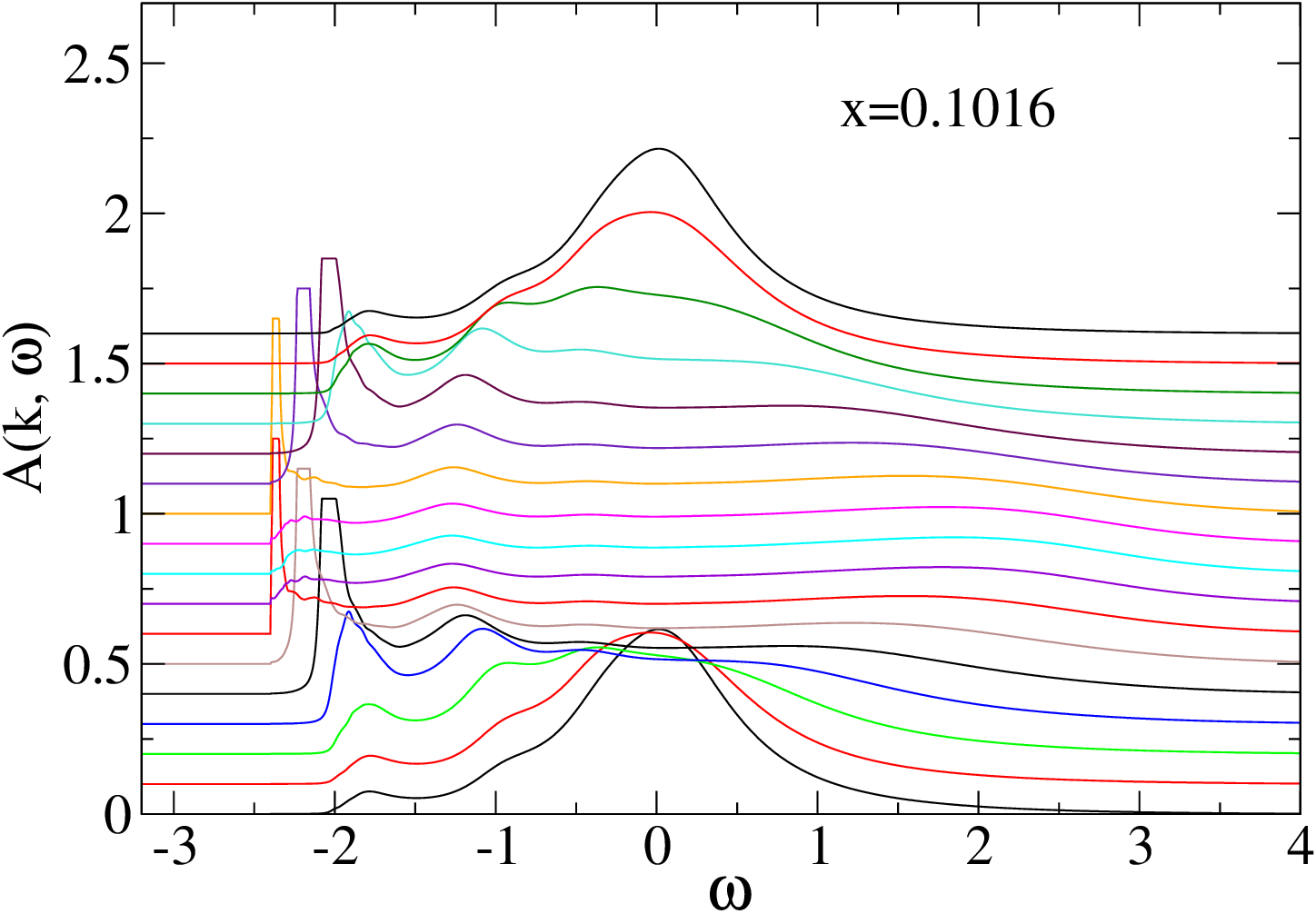,width=0.5\linewidth,clip=}
\end{tabular}
\end{center}
\caption{Hole  spectral  function of the 2D $t-t'-t''-J$ model 
at $J=0.4t (L=32)$ for various doping concentrations $x$.
The wavevector is
varied along the diagonal $(0,0) \to (\pi,\pi)$ direction of the Brillouin
zone. }
\label{figd3}
\end{figure}

The  hole spectral functions for various doping concentrations
are  shown in  Fig.~\ref{figd2}  and Fig.~\ref{figd3} (in all of
our figures where we present spectral functions we do that for $\omega$ 
greater than the Fermi level) for  the cases of the $t-J$  and
$t-t^\prime-t^{\prime\prime}-J$  model   respectively.  
The various lines in a given graph correspond to different wavevectors
which  are
varied along the diagonal $(0,0) \to (\pi,\pi)$ direction of the Brillouin
zone with the lowest curve corresponding to ${\bf k}=(0,0)$ and the
middle to ${\bf k}=(\pi/2,\pi/2)$. 
There is a well-defined
quasiparticle peak at $(\pi/2,\pi/2)$ which we have truncated at some 
finite value in order to make the other features of the spectra visible.
We can  see that with increasing doping the  quasi-particle peak at 
$(\pm \pi/2,\pm\pi/2)$
gets truncated  at the  Fermi level (as it falls below the
Fermi level upon low doping)  and  the higher  energy string  states
gradually  lose their strength and become broader. 
Fig.~\ref{figd2} also illustrates 
that for ${\bf k}=(0,0)$ the lowest energy quasiparticle fades away
with doping and the higher energy ``string'' states accumulate
most of the spectral weight as the doping increases. Namely, starting
from the quasiparticle peak at $(\pi/2,\pi/2)$ and moving towards 
$(0,0)$, we see a gradual transfer of the spectral weight from the
lowest energy peak to the higher energy ``string'' states. 
This is exactly what 
was proposed in Ref.~\onlinecite{strings} as the origin of the waterfall-like 
features. Fig.~\ref{figd3} illustrates that a very similar behavior holds 
for the $t-t'-t''-J$ model.

\begin{figure}[htp]
\begin{center}
\vskip .3 in
\epsfig{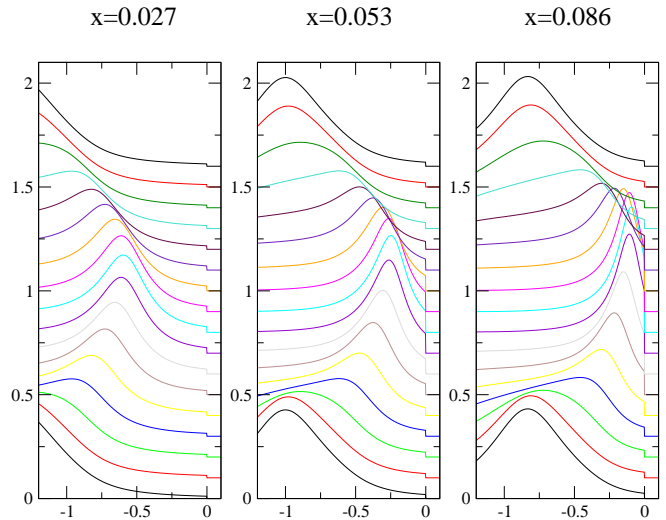}
\end{center}
\caption{Hole  spectral  function of the 2D $t-t'-t''-J$ model 
at $J=0.4t$  for various doping concentrations $x$ calculated 
with a large value of $\eta=0.4t$.}
\label{width}
\end{figure}

\begin{figure}[htp]
\vskip .3 in
\begin{center}
\epsfig{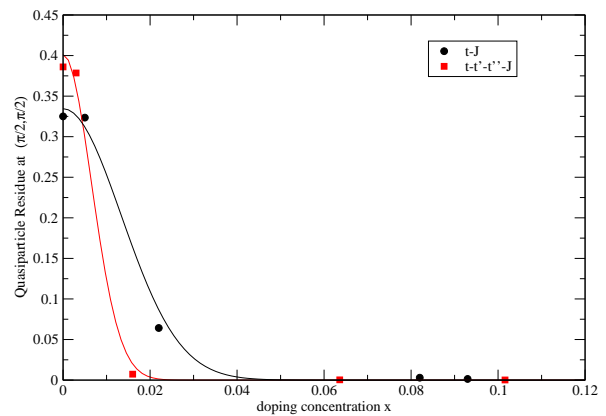}
\end{center}
\caption{Quasiparticle residue   for  ${\bf k}=(\pi/2,\pi/2)$  as a function
of hole concentration for the $t-J$ and the $t-t^\prime-t^{\prime\prime}-J$ model
  at $J=0.4t$.}
\label{residue}
\end{figure}

Sometime ago Wells 
et al.\cite{wells} found that there is a
well-defined quasiparticle peak at $(\pi/2,\pi/2)$,
within their low resolution ARPES data, 
which follows a dispersion with a bandwidth which is also in agreement with the
results for the $t-J$ and related models\cite{liu,strings}.
Later, however, improved resolution ARPES measurements by
Shen et al.\cite{shen2} reveal that this low energy peak is significantly
broader than the limit imposed by the experimental resolution. 
In addition, Shen et al.\cite{shen2}
find that the chemical potential, at low doping concentrations
$x$, is significantly shifted from the lowest energy quasiparticle
peak, thus, appearing to enter the Mott gap.
Furthermore, rather recently Fournier et al.\cite{fournier} 
find that the quasiparticle
residue is essentially zero below doping concentration of $0.1$. 
Next we will try to understand these results within the results
of our calculation. First of all, our present calculation which ignores
the role of phonons shows that the lowest energy quasiparticle peak
at $\vec G$ is much sharper than the experimental peak. 
Previously\cite{satyaki}, we added the role of optical phonons 
and of finite temperature, both of which have been neglected in the present
calculation,  and 
we demonstrated that the lowest energy peak becomes broadened.
In Fig.~\ref{width} we present the results of our calculation
obtained using the $t-t'-t''-J$ model for three different values
of doping $x$ and a relatively large value of the broadening parameter 
$\eta$, in order to compare with Fig.~3 of Ref.~\onlinecite{shen2}.
Notice that the evolution of spectra seem qualitatively very similar
to those reported in Ref.~\onlinecite{shen2}. In particular, we notice
that the broadening causes the position of the chemical potential
(the value of $\omega$ in the spectra in Fig.~\ref{width} has been 
shifted by the value of $\mu$ such that the position of $\mu$ is
at $\omega=0$) to move significantly away from the quasiparticle peak
inside the Mott gap. This is very similar to the observation\cite{shen2}. 
Therefore, we find that the results
reported in Ref.~\onlinecite{shen2}, regarding what was referred to as
``paradoxical shift'' of the chemical potential, can be understood as a
 consequence of the broadening.

Fig.~\ref{residue} illustrates the quasiparticle residue as a function
of doping for the $t-J$ and the $t-t'-t''-J$ model for $J/t=0.4$ obtained
by integrating the spectral function from $\omega=\mu$ up to the first 
minimum of the spectral function for $ \omega > \mu$.  Our results on 
the $t-t'-t''-J$ model
indicate that the quasiparticle residue becomes very small away from the undoped
limit and as  can be inferred from Fig.~4
of Ref.~\onlinecite{fournier} our results are within error bars 
of the experimental
values. In addition, Fournier et al.\cite{fournier} provide no data in the 
immediate vicinity of zero doping.
  
Therefore, we can conclude that our results can explain the 
ARPES\cite{wells,shen2,fournier} results without additional 
hypotheses.

\begin{figure}
\includegraphics[width=\linewidth]{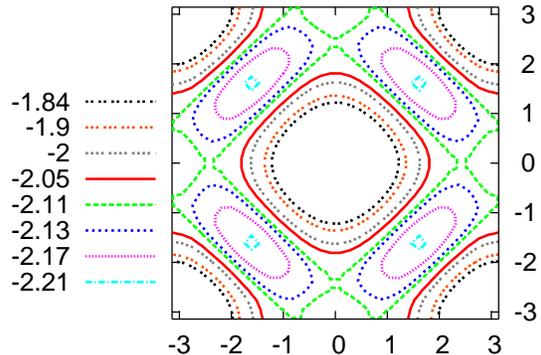}
\caption{Contour plots of the rigid hole energy dispersion
$E(k_x,k_y)$  in  the $t-J$ models for $J=0.4t$ in a
  $32 \times 32$ lattice .}
\label{figd8}
\end{figure}

\begin{figure}
\begin{tabular}{c}
\epsfig{file=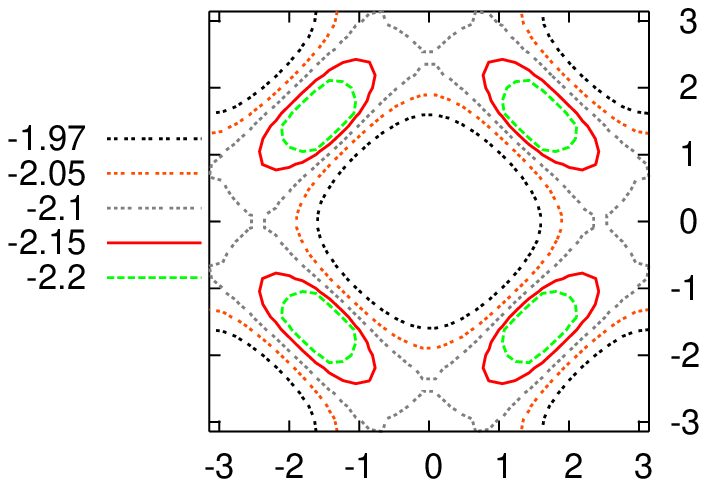,width=.9\linewidth,clip=}\\
\epsfig{file=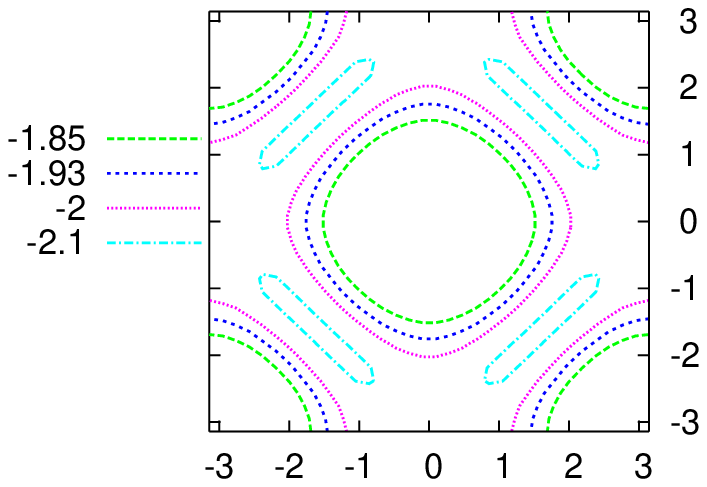,width=.9\linewidth,clip=}\\
\epsfig{file=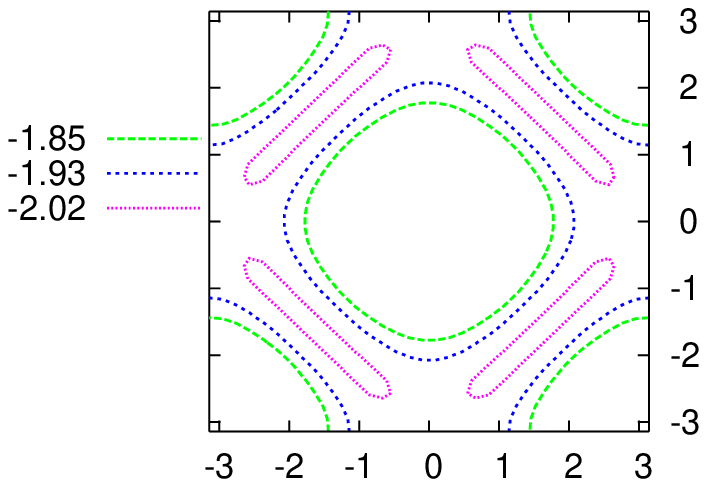,width=.9\linewidth,clip=}
\end{tabular}
\caption{Contour plots of the hole energy dispersion
$E(k_x,k_y)$  for  the $t-J$ model  for $J=0.4t$ in  a $32
  \times    32$   lattice.    Top:    $\mu=-2.20t,~x=0.022$;   middle:
  $\mu=-2.10t,~x=0.082$; bottom: $\mu=-2.02t,~x=0.118$.}
\label{figd6}
\end{figure}

\begin{figure}
\begin{tabular}{cc}
\epsfig{file=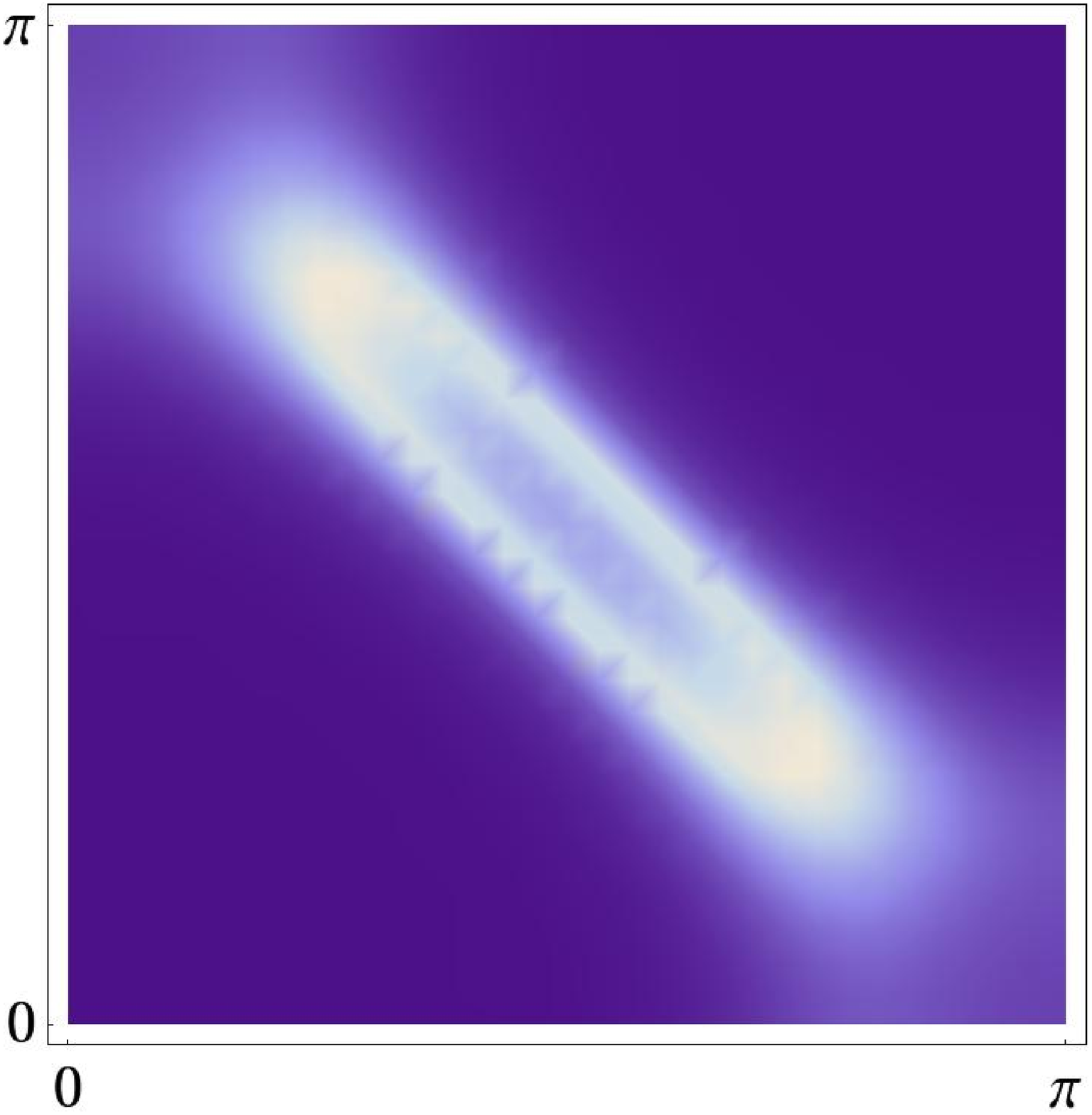,width=.45\linewidth,clip=}&
\epsfig{file=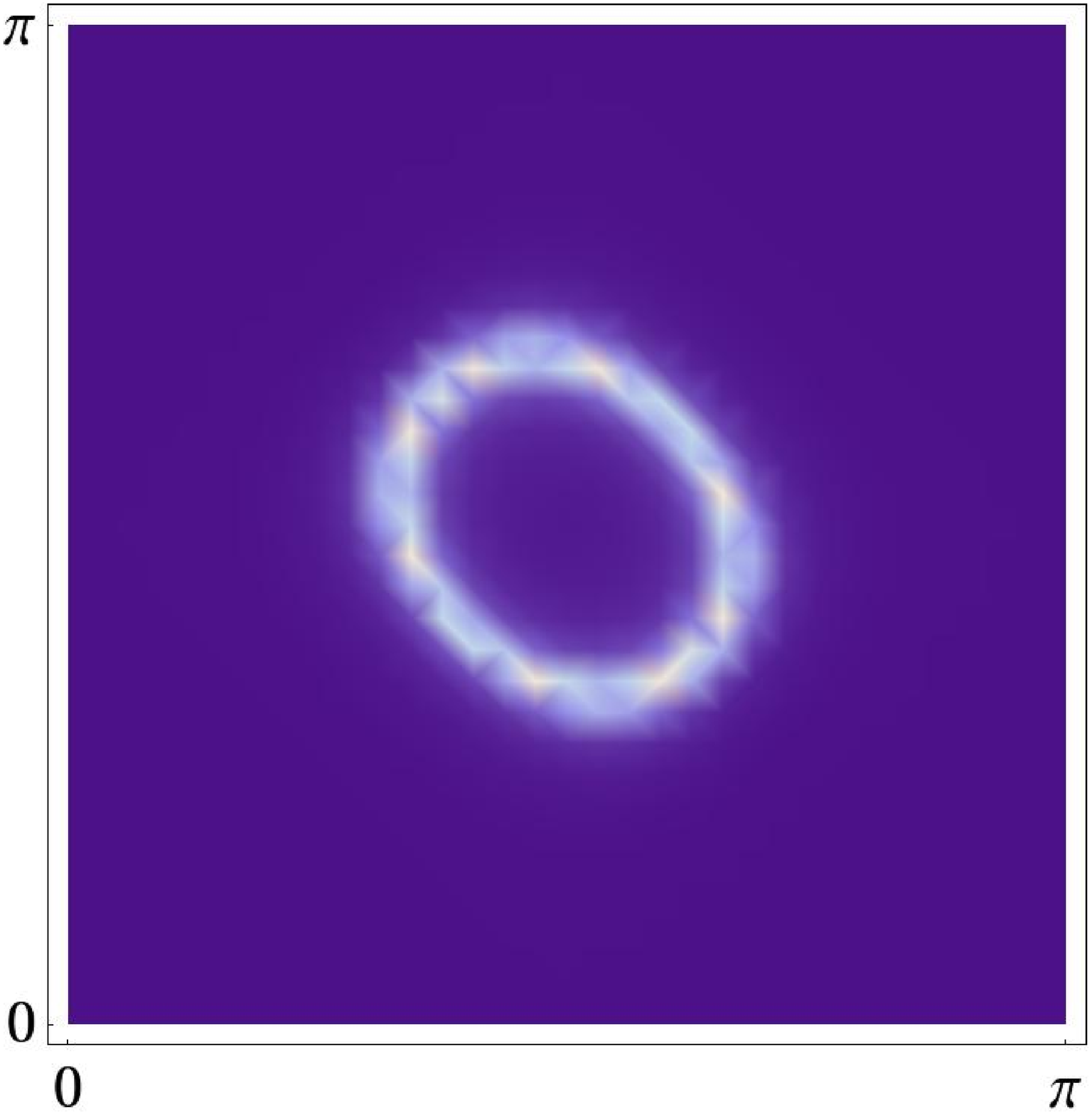,width=.45\linewidth,clip=}\\
\end{tabular}
\hskip 1. in \epsfig{file=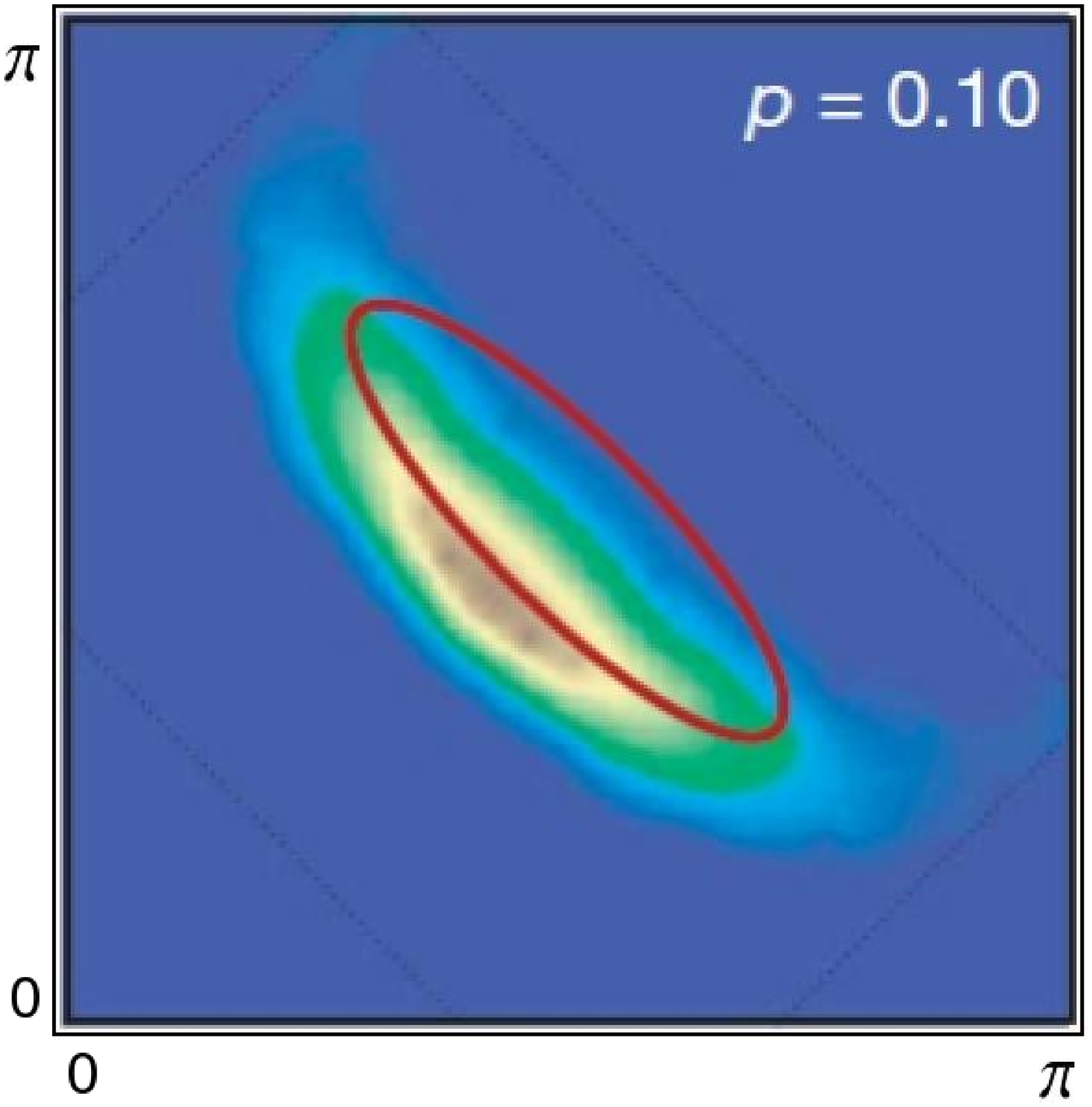,width=.45\linewidth,clip=}
\caption{Intensity  plots of hole spectral functions at  the Fermi energy
  for $J=0.4t$  as calculated on a  $64 \times 64$ lattice  
for  the  $t-J$ model (top left) at doping concentration $x=0.084$ and for the
$t-t'-t''-J$ model (top right)  at doping concentration      $x=0.094$.      
 Bottom:     ARPES  intensity   plot  of
  $Na_{2-x}Ca_xCu_2O_2Cl_2$ at  $x=0.1$ done in 
 Ref.~\onlinecite{shen} (this part of the figure is taken from 
Ref.~\onlinecite{doiron} which was reproduced from Ref.~\onlinecite{shen}
and the red elliptical Fermi arc was added).  All plots show  only the 1st
  quadrant of the Brillouin zone.}
\label{fig18}
\end{figure}

\begin{figure}[htp]
\vskip .2 in
\begin{center}
\includegraphics[width=.9\linewidth]{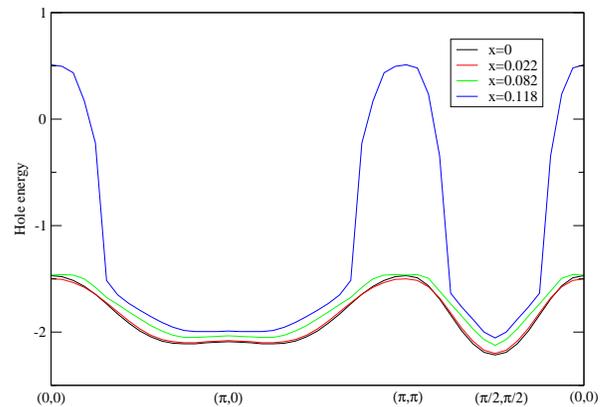}
\end{center}
\caption{Hole energy  band in the $t-J$  model for $J=0.4t$ in  a $ 32
  \times   32$  lattice  along   $(0,0)\rightarrow  (\pi,0)\rightarrow
  (\pi,\pi)\rightarrow(\pi/2,\pi/2)\rightarrow     (0,0)$    cut    in
  $k$-space.}
\label{figd19}
\end{figure}

\subsection{Hole energy dispersions}
\label{dispersions}

 Fig. \ref{figd8}  shows a  2D contour plot  of the rigid  hole energy
 dispersion  $E(k_x,k_y)$ derived from  the $t-J$  model on  a $32  \times 32$
 lattice for  $J=0.4t$. As we follow  the iso-energy lines  we can see
 that  the  low  energy  contours  are  ellipses  around  the wave  vectors
 $(\pm\pi/2,\pm\pi/2)$   (the   lowest    energies   are   indeed   at
 $(\pm\pi/2,\pm\pi/2)$  which is  not shown  in the  plot).  At higher
 energy, the  topology gradually changes and the  ellipses increase in
 size and  become connected to each  other giving rise  to new pockets
 around  $(0,\pm\pi)$  and  $(\pm\pi,0)$.  This transition  occurs  at
 doping around $x\sim$0.145. At  even higher energy these newly formed
 pockets  disappear and  disconnected contours  are  obtained centered
 around (0,0) and $(\pm\pi,\pm\pi)$  and happens above $x\sim0.19$.  A
 similar  change  of  topology  as  function  of  $x$  occurs  in  the
 $t-t'-t''-J$  model; the  main  difference being  that the  $k$-space
 anisotropy  of the  spectral  intensity is  reduced  making the  hole
 pockets more circular than elongated.

The contour  plot of the  energy dispersion obtained for  doped system
show  similar features  at  low doping. 
The hole energy dispersion
$E(k_x,k_y)$ is obtained as the lowest energy peak above $\mu$ in the spectral 
function as a function of ${\bf k}$. Fig.  \ref{figd6} shows  the
contour plots  for hole  energy dispersions at  various dopings. We  can see
that  the   Fermi  surfaces  that   are  small  hole   pockets  around
$(\pm\pi/2,\pm\pi/2)$  for  very  small  doped system,  gradually  get
elongated along anti-nodal directions as the system is doped with more
and more  holes. 

In Fig. \ref{fig18},  the calculated hole  spectral 
intensity plots
at  Fermi energy,  for the $t-J$  model (for  $x=0.084$, top left), and  for
the $t-t'-t''-J$ (for $x=0.094$, top right) are compared 
with the  experimental ARPES plot (bottom) for  the     $Na_{2-x}Ca_xCu_2O_2Cl_2$
sample with $x=0.1$ taken from Ref.~\onlinecite{shen}. 
 The experimental   intensity plot has
a banana-like shape close  to $(\pm\pi/2,\pm\pi/2)$  at this doping.  
Our numerical
results  show elliptical rings centered about  
$(\pm\pi/2,\pm\pi/2)$  
stretched  along  off-diagonal  direction  for the  $t-J$  model.  
Our calculations are restricted in the regime where there is
long-range antiferromagnetic order present in the system where the hole
spectra should be symmetric against reflections  with respect to the
$(0,\pi) \to (\pi,0)$ line in the Brillouin zone. As a result spectra
having banana-like shape cannot exist in the $t-J$ or $t-t'-t''-J$ model 
when there is AF order. 
It is possible when the long-range antiferromagnetic order
is destroyed upon further doping, where the reflection symmetry 
with respect to the $(0,\pi) \to (\pi,0)$ line in the Brillouin zone
is broken, that the ring-like shape of the
spectral intensity plot might transform to  half elliptical ring-like
shape which might resemble the banana shape.
For example, if we cut the intensity distribution obtained for 
the $t-t'-t''-J$  model shown in Fig. \ref{fig18} (top right) along
the $(0,\pi) \to (\pi,0)$ line, one of the pieces will resemble the 
experimental intensity plot.  The next  nearest
neighbor  hoppings present in the $t-t'-t''-J$ model  reduce  the   
$k$-space  anisotropy  of  the  hole
pockets.    However, the conclusion drawn from quantum oscillation 
measurements\cite{doiron} indicate that there may be elliptical 
pockets centered around
the $(\pm\pi/2,\pm\pi/2)$  points. These are shown by the ellipse in 
Fig.~\ref{fig18} (bottom). 
Therefore, it is not clear why only one
side of the elliptical Fermi arcs are luminous in the ARPES studies.
The ARPES result indicates that the quasiparticles are much better
defined only just outside of one side of the Fermi arcs.

We should mention  here that  because of  the  finite 
energy resolution  $\eta$ in  our
calculation, in  Fig. \ref{fig18} we  have also smoothed
the spectral functions as follows
\begin{eqnarray}
A(k,\omega)=\int                      A(k,\omega^\prime)\frac{\eta}{\pi
  ((\omega-\omega^\prime)^2+\eta^2)}d\omega^\prime
\label{smoothing}
\end{eqnarray}
with $\eta=0.02t$.

Fig.  \ref{figd19}  shows  the  hole  bands for the  $t-J$  model  along
$(0,0)\rightarrow                                    (\pi,0)\rightarrow
(\pi,\pi)\rightarrow(\pi/2,\pi/2)\rightarrow   (0,0)$   direction   in
$k$-space. The reason that a considerably larger bandwidth of the order  
of $t$ is found  at   higher    doping   ($x=0.118$)    (also    shown   in
Ref. \onlinecite{sherman}) is that the lowest energy string state, as 
${\bf k}$ approaches $(0,0)$ or $(\pi,\pi)$,
loses  all of its spectral weight which is transferred to higher energy
string-states.  However, the  staggered  magnetization  ($M_s$)  is 
zero  at  those  doping  concentrations as  can  be  seen  from
Fig. \ref{figd13} and so our calculation of the hole band 
for this value of $x$ is outside the range of validity of our calculation.

\subsection{Waterfall-like features}
\label{waterfall}

\begin{figure}
\begin{tabular}{cccc}
\epsfig{file=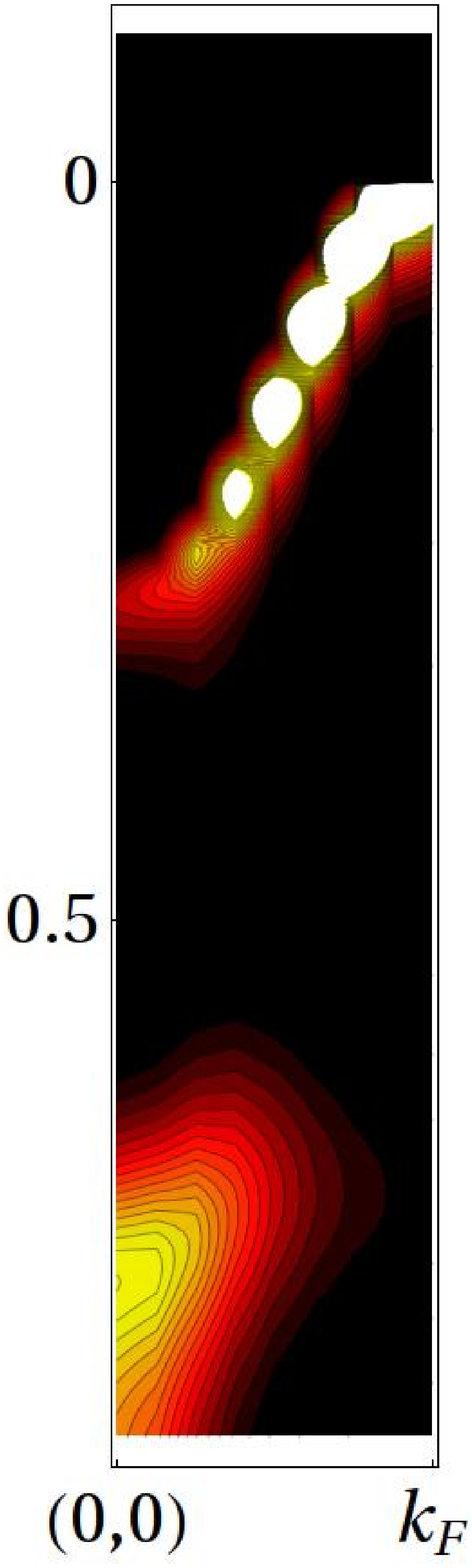,width=.25\linewidth,clip=}&
\epsfig{file=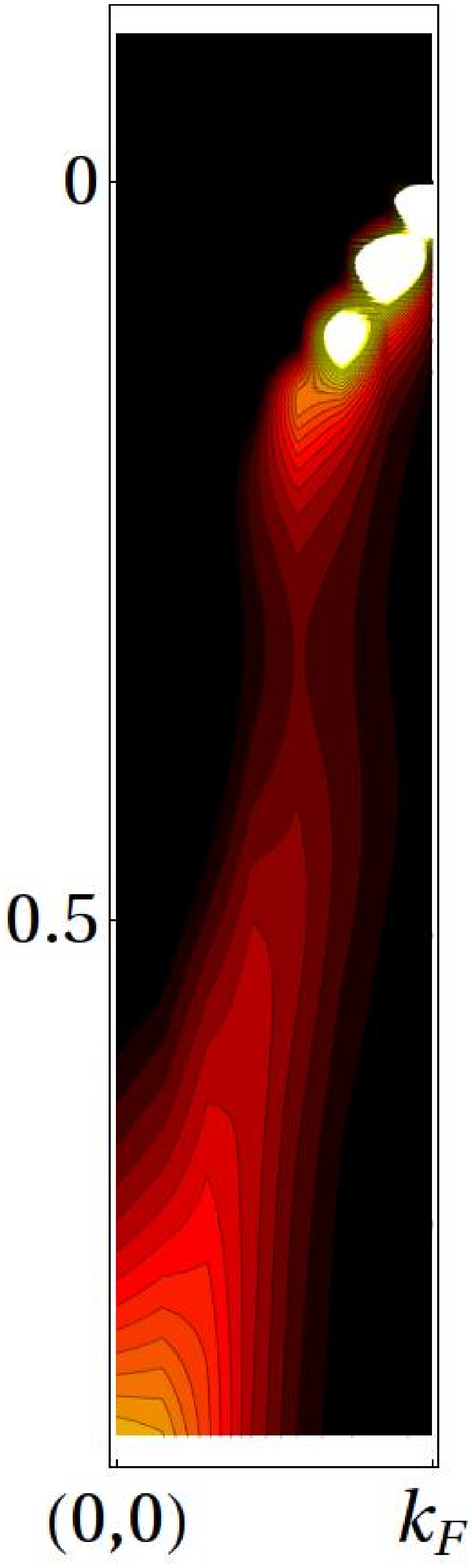,width=.25\linewidth,clip=}&
\epsfig{file=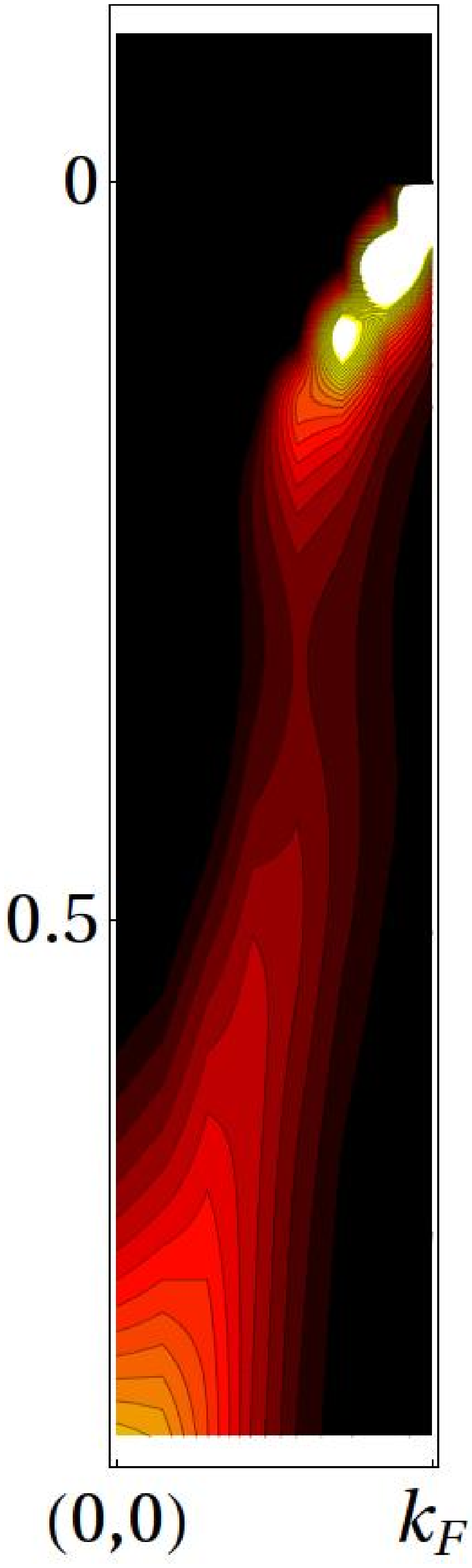,width=.25\linewidth,clip=}&
\epsfig{file=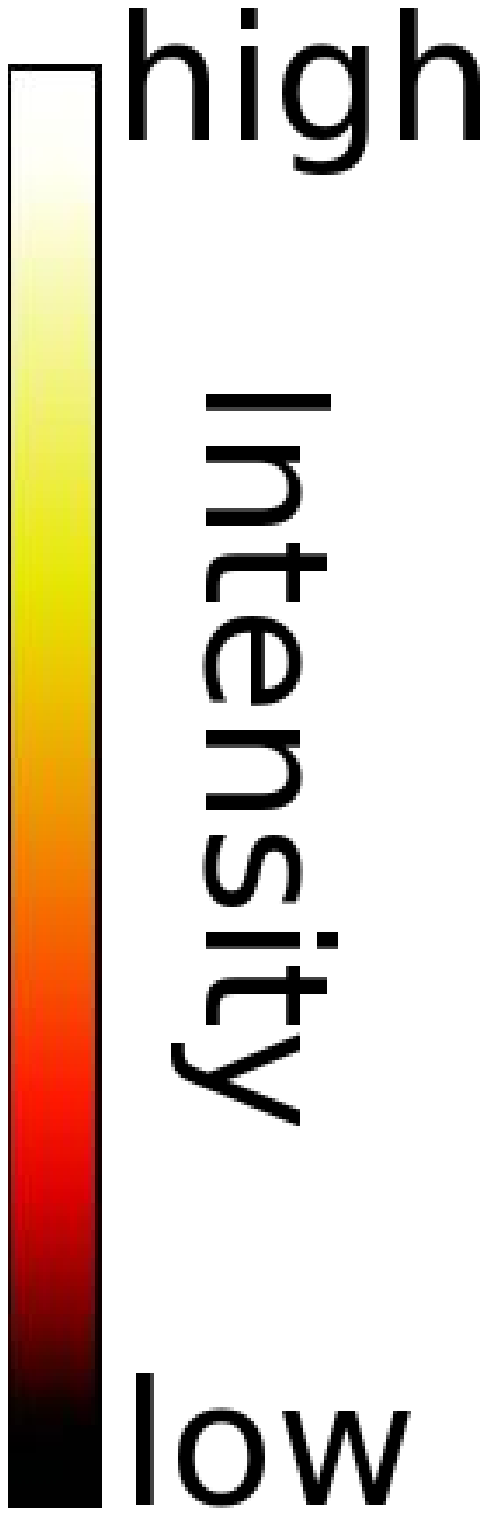,width=.09\linewidth,clip=}
\end{tabular}
\caption{Intensity plots in  a $32 \times 32$ lattice  from a 2D $t-J$
  model  for   doping  concentration  $x=0.02$   (left)  and  $x=0.09$
  (middle,right) for $J=0.4t$. Lorentzian broadening with $\eta=0.03t$
  is used in plot at right.}
\label{watfl}
\end{figure}
\begin{figure}
\begin{tabular}{ccccc}
\epsfig{file=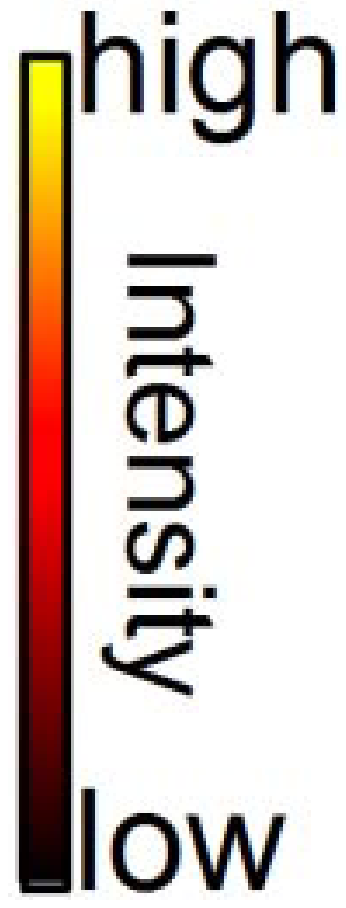,width=.075\linewidth,clip=}&
\epsfig{file=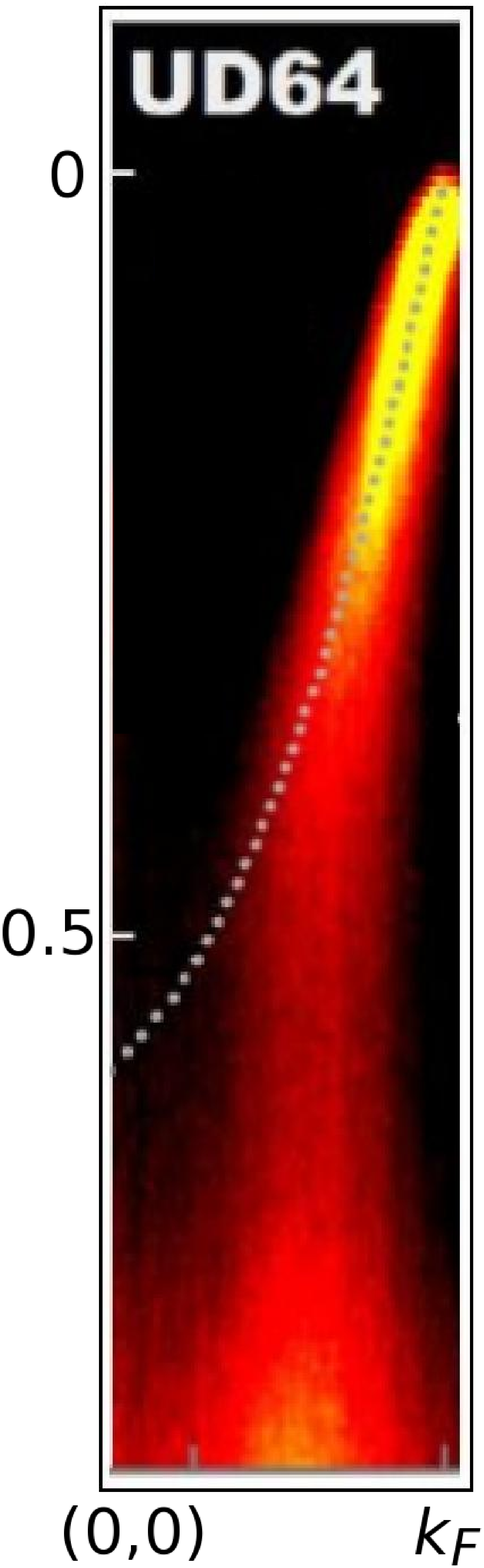,width=.26\linewidth,clip=}&
\epsfig{file=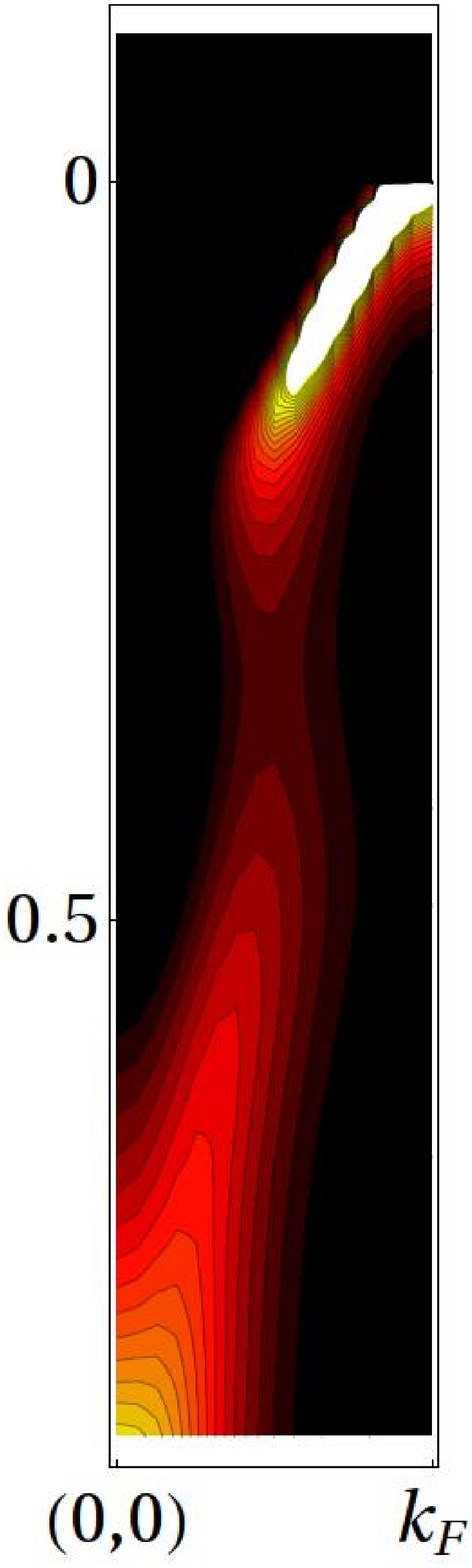,width=.25\linewidth,clip=}&
\epsfig{file=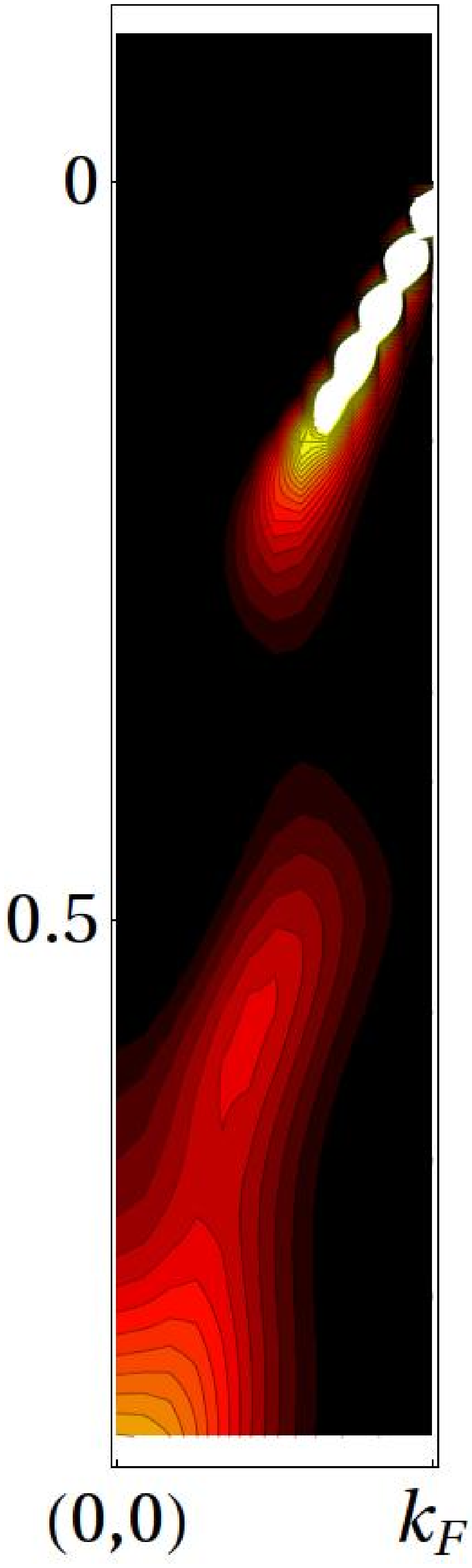,width=.25\linewidth,clip=}&
\epsfig{file=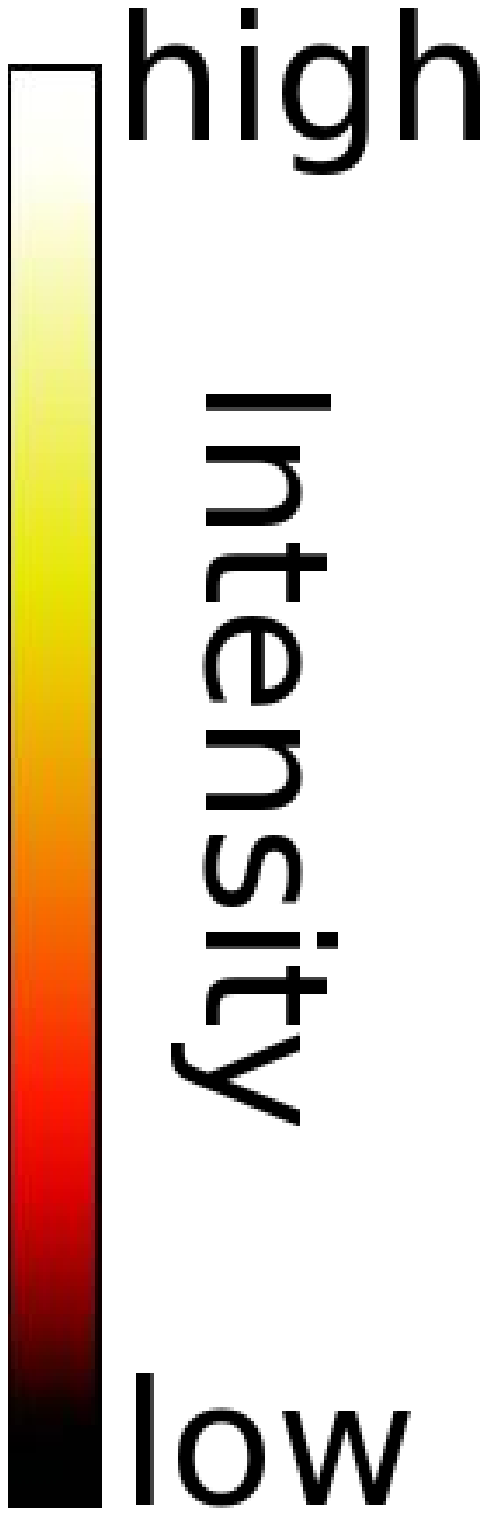,width=.09\linewidth,clip=}
\end{tabular}
\caption{Waterfall  feature:  ARPES  intensity  plot  \cite{graf}  of
  under-doped  Bi2212 (left) and  intensity plots  in a  $64 \times 64$ lattice
  from a  2D $t-J$ model  for doping concentration  $x=0.084$ (middle)
  and from a $t-t'-t''-J$ model for $x=0.094$ (right) for $J=0.4t$.}
\label{figarpes}
\end{figure}

 The  single-hole spectral function as obtained from the
$t-J$  model has  high energy  string states \cite{liu}  and these
states,  along with  the  quasi-particle peak,   broaden in  the
presence    of     optical    phonons    and     finite    temperature
\cite{satyaki}. This enables a smooth transition of intensity from the
low energy  peak at $(\pi/2,\pi/2)$ to  the high energy  peak at (0,0)
\cite{satyaki}  as  seen  in  the  experimental  waterfall-like  feature
\cite{damascelli}.  Here, we find  a similar  fading  away of  the higher
energy string states  with doping  and that makes  a bridge between  the high
energy and  low energy states so  that the continuous change in the intensity 
gives rise to a  waterfall-like pattern. 

The spectral
functions are  smoothed as discussed by means of Eq. \ref{smoothing} 
to smooth out the ``sharp edges'' produced by the combination of the
finite-size system and the 
finiteness  of the  energy resolution  used in the  numerical calculations.

Fig.  \ref{watfl} shows  intensity plots  along the nodal  direction obtained
using the $t-J$  model for  $x=0$ and  $x=0.09$.  The discreteness  of the  high
intensity close to Fermi wave-vector $k_F$ is due to finite size effects and
can be    smoothed    out   when using   higher   values    of    $\eta$. 
Fig. \ref{figarpes} compares  intensity plots obtained from the $t-J$ and
the $t-t'-t''-J$  model  for a $64 \times 64$  size  lattice  with the  ARPES
waterfall-like  intensity plots   on the under-doped
Bi2212. Notice a  smooth transfer of  intensity in the  $t-J$ model
intensity plots resembling the experimental ARPES plot \cite{graf}.

\section{Magnetic order}
\label{magnetic}
\subsection{Magnon spectral functions}
\label{magnons}
\begin{figure}[htp]
\begin{center}
\begin{tabular}{c}
\epsfig{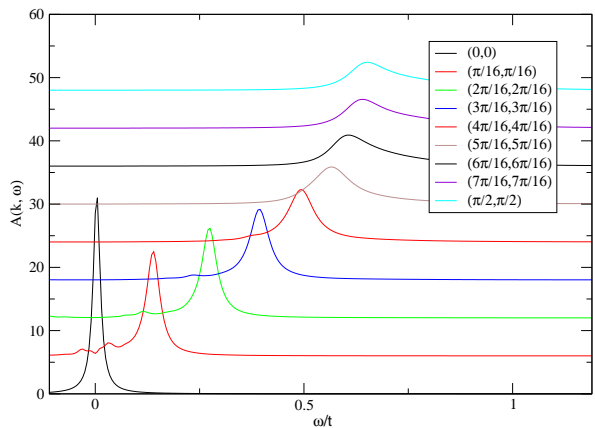}
\end{tabular}
\end{center}
\caption{Magnon            spectral            functions           for
along the diagonal of the Brillouin zone
obtained for the
  $t-J$ model for $J=0.4t$ (L=32) for $x=0.022$.}
\label{fig9a}
\end{figure}
\begin{figure}[htp]
\begin{center}
\begin{tabular}{c}
\epsfig{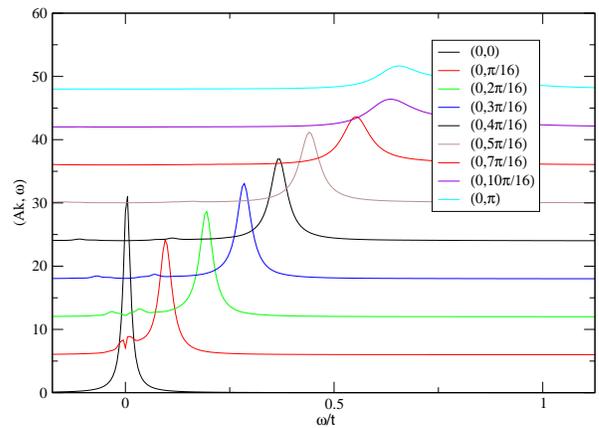}
\end{tabular}
\end{center}
\caption{Magnon            spectral            functions           for
along the $(0,1)$ direction of the Brillouin zone
obtained for the
  $t-J$ model for $J=0.4t$ (L=32) for $x=0.022$.}
\label{fig9b}
\end{figure}
\begin{figure}[htp]
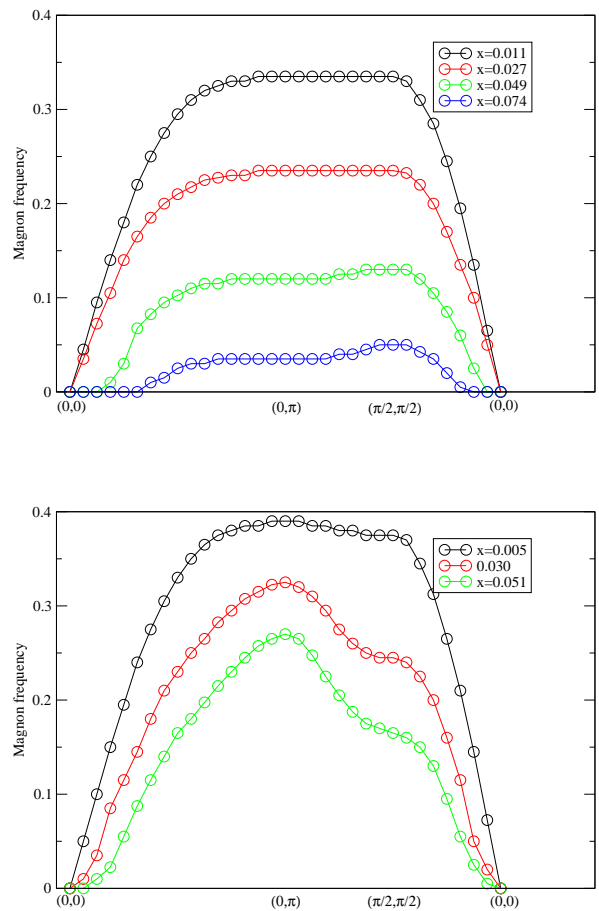

\begin{center}
\includegraphics[width=.9\linewidth]{Fig14a.eps}\vspace{.45
  in}\\ \includegraphics[width=.9\linewidth]{Fig14b.eps}
\end{center}
\caption{Magnon    frequency     in      the     $t-J$     (top)     and
  $t-t^\prime-t^{\prime\prime}-J$ (bottom) models for $J=0.2t$.}
\label{figd10}
\end{figure}

In Fig.~\ref{fig9a} and Fig.~\ref{fig9b} we present the magnon spectral 
function for $x=0.022$ 
(for a $32\times 32$ size lattice) of the $t-J$ model along the diagonal and
the $(1,0)$ directions of the Brillouin zones respectively. 
For both the undoped and doped system,
 $D(k,\omega)$ has a rather featureless peak  at the renormalized magnon
energy $\Omega_r(k)$  with a width of similar 
order of magnitude to $\Omega_r(k)$ which is of order $J$, unlike  
the hole spectral
function, which has a spread from approximately -4t to 5t
in energy. The  renormalized magnon energy 
decreases rapidly with increased doping 
and  at some low value of doping concentration $x$ the spin  wave spectrum  
 collapses  at small wave-vectors and this signals that the AF-LRO disappears
(see also Ref. \onlinecite{Ko}).  
 Fig. \ref{figd10} shows the spin
wave  dispersion as a function of doping 
obtained from  the  $t-J$ and  $t-t^\prime-t^{\prime\prime}-J$
models  in a  $32 \times  32$ lattice.  Doping  lowers the magnon frequency
and above a certain value of $x$ it becomes  zero starting  at 
low  wave-vectors.  

\begin{figure}[htp]
\begin{center}
\begin{tabular}{c}
\epsfig{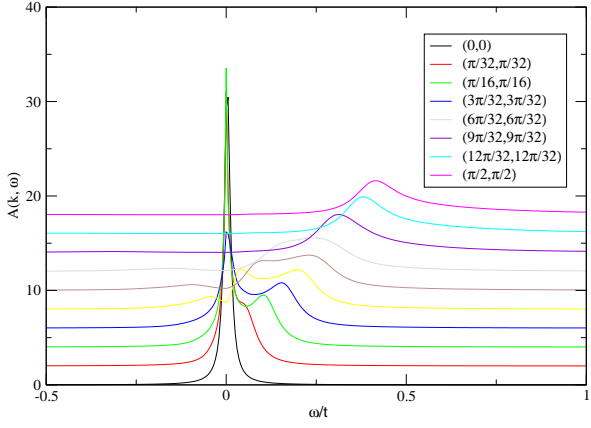}
\end{tabular}
\end{center}
\caption{Magnon            spectral            functions           for
along the diagonal of the Brillouin zone
obtained for the
  $t-J$ model for $J=0.4t$ (L=64) for $x=0.084$.}
\label{fig11a}
\end{figure}
\begin{figure}[htp]
\begin{center}
\begin{tabular}{c}
\epsfig{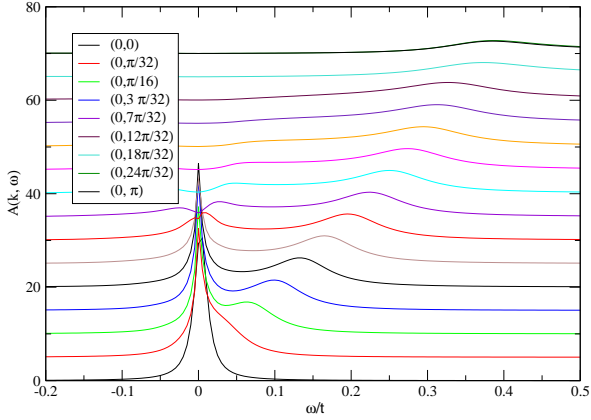}
\end{tabular}
\end{center}
\caption{Magnon            spectral            functions           for
along the $(0,1)$ direction of the Brillouin zone
obtained for the
  $t-J$ model for $J=0.4t$ (L=64) for $x=0.084$.}
\label{fig11b}
\end{figure}

Fig.~\ref{fig11a} and Fig.~\ref{fig11b} shows the magnon spectral function 
for $x=0.084$ 
(for a $64\times 64$ size lattice) of the $t-J$ model along the diagonal and
the $(1,0)$ directions of the Brillouin zones respectively.  Notice that
in this case however, along with the broad magnon peak at the 
renormalized magnon frequency there is a sharp peak at $\omega =0$
for small wavevectors.
This is an indication of  an    instability   
which was also  found  by    Krier   et al. \cite{krier}. 
This instability could be due to the spiral 
order\cite{shraiman,jayaprakash,yoshioka} or due to stripe 
ordering\cite{white}. 

\subsection{Staggered magnetization}

\label{smagnetization}
\begin{figure}[htp]
\vskip .3 in
\begin{center}
\includegraphics[width=.9\linewidth]{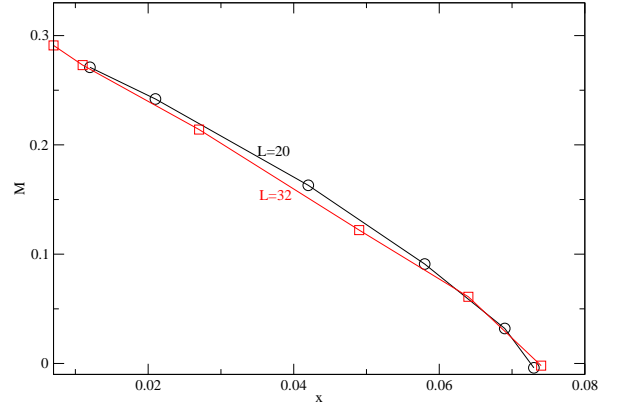}
\end{center}
\caption{Staggered Magnetization  in the $t-J$  model for L=20  and 32
  ($J=0.2t$).}
\label{figd11}
\end{figure}
\begin{figure}[htp]
\vskip .3 in
\begin{center}
\includegraphics[width=.9\linewidth]{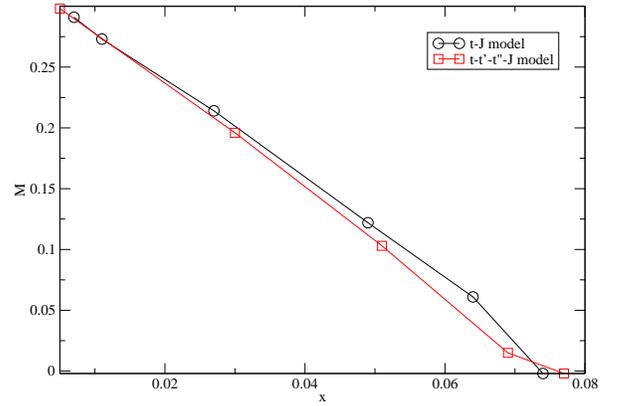}
\end{center}
\caption{Staggered     Magnetization      in     the     $t-J$     and
  $t-t^\prime-t^{\prime\prime}-J$ models for $J=0.2t$ (L=32).}
\label{figd12}
\end{figure}
\begin{figure}[htp]
\vskip .2 in
\begin{center}
\includegraphics[width=.9\linewidth]{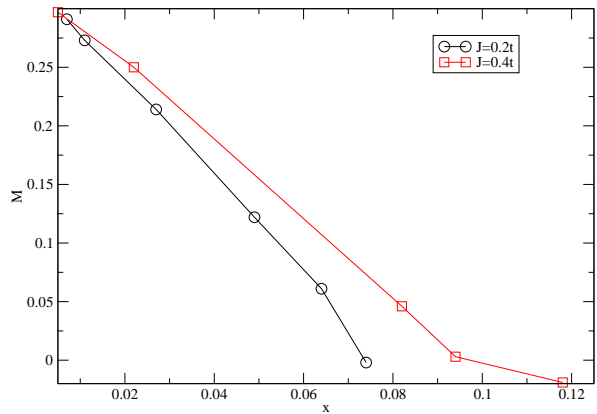}
\end{center}
\caption{Staggered Magnetization  in the $t-J$ model  for $J=0.2t$ and
  $0.4t$ (L=32).}
\label{figd13}
\end{figure} 

The    staggered   magnetization   ($M_S$)    for   a    spin-1/2   2D
antiferromagnetic  square  lattice  is  0.303  \cite{RMP}.  With  hole
doping,  $M_S$  decreases  from the above  value. Fig.~\ref{figd11}
indicates that the role of finite-size effects may be insignificant. 
We find that 
$M_S$ decreases steadily with doping and eventually goes to zero within
a doping  range between 0.07  and 0.08 for  the value of  $J=0.2t$ (as
seen in  Fig. \ref{figd11} and  \ref{figd12}). Using larger  values of
$J$ increases  the value  of the critical  doping ($x_c$)  where $M_S$
ceases  to  exist.  Fig.  \ref{figd13}  shows that  for  $J=0.4t$  the
magnetization goes to zero at  a concentration $x_c\sim$  0.12. Our
staggered  magnetization  plots  are   similar  to  that  obtained  in
Ref. \onlinecite{vojta,horsch,gan,orbach2,belkasri}. 
Following the discussion regarding the possible instabilities of our
previous section, we would like to comment that our calculation 
cannot definitely distinguish the case where the antiferromagnetic
order disappears due to the presence of the stripe or spiral or
phase separation instabilities\cite{hellberg2} at any doping
from the case where the order disappears at a finite concentration.
However, the fact that the staggered magnetization feels weak
finite-size effects is an indication of the latter scenario.

\section{Conclusions}
\label{conclusions}

We  have studied the  2D $t-J$  model and  the
related $t-t'-t''-J$ model in the low doping regime where the antiferromagnetic
LRO is present. Our approach requires the existence of AF-LRO 
in order to be qualitatively correct because it starts by transforming these 
Hamiltonians by keeping up to quadratic terms of operators which create 
spin-deviations above the N\'eel state. We find that doping
quickly reduces the degree of magnetic  order and 
completely destroys it for doping concentration greater than
$x\sim  0.08 -  0.1$ (for $0.2 \le J/t\le 0.4$).  
Experimental findings on cuprate superconductors also  show quick  
disappearance of  long
range  magnetic  order   with  doping  \cite{keimer} and  when the LRO is
destroyed by doping the correlation length is several  lattice constants. 

We find that the residue of the quasiparticle peak at $(\pm\pi/2,\pm \pi/2)$
dies out quickly with doping and in particular in the $t-t'-t''-J$ model 
it becomes negligible away from the undoped limit. In addition, we 
find that when we  broaden the lowest energy quasiparticle peak, the
spectral function shares most of its features with those obtained by
high resolution ARPES studies\cite{shen2}. In particular, we find that
what is called  in Ref.~\onlinecite{shen2} ``paradoxical shift''  
of the chemical potential within the Mott gap, is a natural result
of the broadening of the lowest energy quasiparticle peak.
 Our results indicate that the high resolution
ARPES results obtained by Shen et al.\cite{shen2} and those
recently reported by Fournier et al\cite{fournier} in the low doping regime 
can be understood in a rather simple way.

Furthermore, we studied the dependence on hole doping 
of the hole pockets which start
forming at $(\pm \pi/2,\pm\pi/2)$ at very low hole doping when antiferromagnetic
order is present in the system. We find that 
as we increase doping
 the  Fermi ``surface'' starts as four  ellipses  around  the wave  vectors
 $(\pm\pi/2,\pm\pi/2)$ and the  topology gradually changes where
 the  ellipses increase in
 size and  become connected to each  other giving rise  to new pockets
 around  $(0,\pm\pi)$  and  $(\pm\pi,0)$.   At  even somewhat higher doping 
these newly formed  pockets  disappear and  disconnected parts of 
the Fermi surface   are  obtained centered
 around (0,0) and $(\pm\pi,\pm\pi)$.  A
 similar  change  of  topology  as  function  of  $x$  occurs  in  the
 $t-t'-t''-J$  model, the  main  difference being  that the  $k$-space
 anisotropy  of the  spectral  intensity is  reduced  making the  hole
 pockets more circular than elongated. Similar pockets have been
inferred from quantum oscillation measurements\cite{doiron}
at relatively low doping.

We also find that the intensity plot of the spectral intensity 
has features with a ring-like shape which are centered around 
the wave vectors $(\pm\pi/2,\pm\pi/2)$.
Our calculations are restricted in the regime where there is
long-range antiferromagnetic order present in the system where the hole
spectra should be symmetric against reflections with respect to the
$(0,\pi) \to (\pi,0)$ line in the Brillouin zone. As a result spectra
having banana-like shape as seen experimentally\cite{shen}, 
cannot exist within the models studied here when there is AF order.
We argued, however, that it is possible when the long-range 
antiferromagnetic order
is destroyed upon further doping that these features with the
ring-like shape of the spectral intensity plot 
could transform to a half ring-like
shape (i.e., a ring centered at $(\pi/2,\pi/2)$ which is scissored along the
$(0,\pi) \to (\pi,0)$ line) which might resemble the 
banana shape pockets near
$(\pi/2,\pi/2)$ seen in the experiment\cite{shen}.

We have also examined the role of doping in producing the
waterfall-like features in the spectral function as seen in ARPES. 
In particular, it was suggested in Ref.~\onlinecite{strings} that
these waterfall-like features is the result of spectral weight transfer
from the lowest energy quasiparticle peak to the higher ``string''
states as the momentum varies from $(\pi/2,\pi/2)$ to
$(0,0)$ along the diagonal of the Brillouin zone.
We find that doping significantly broadens the high energy ``string'' states
and this allows a gradual and
continuous  flow of spectral  weight from  the quasi-particle  peak at
$(\frac{\pi}{2},\frac{\pi}{2})$ to  the high energy peak  at (0,0). 
We also find that doping does not broaden the lowest energy
peak enough to have agreement with the experimental
waterfall-like feature, thus, other hole
decay mechanisms such as decay into phonons or broadening due to finite 
temperature\cite{satyaki} may be needed.

\section{Acknowledgments}
We wish to thank A. Damascelli for useful discussions.

\appendix

\section{Technical part}
\label{details}

For the  numerical calculation, we  always keep the  energy resolution, i.e., 
the finite-frequency step
$\Delta\omega$ (0.005t) less than  $\eta$ (we used $\eta$ between 0.4t
and  0.01t)  so  that  the  Lorentzian feature  of  the  bare  Green's
functions   can   be   represented   properly   as   also  discussed   in
Refs.  \onlinecite{liu,satyaki}.  The  energy  range between  which  we
iterate both the Green's functions are $(-8t,8t)$ as the variance of the
spectra are always well within such limit. We should mention here that
the $\Delta\omega$  used in  Ref. \onlinecite{sherman} that  gives 400
equally  spaced  points in  the  range  -5t to  4t  is  larger than  the
Lorentzian broadening, $\eta=0.015 t$ that they used.

We find that the convergence  of the self-consistent equations is slow
for the  doped antiferromagnet (compared  to the calculations  done in
Ref.  \onlinecite{liu}  for  obtaining  the hole  Green's  functions  from
undoped system). Compared to the $t-t^\prime-t^{\prime\prime}-J$ model
the convergence  is even  slower for  the $t-J$ model  as they  have a
complete  degeneracy in the  bare hole  energy. In  order to  obtain a
quick convergence we update  the spectral functions, in each iteration
step,  by mixing  the  values obtained  from  the last  two
iterations-      a      technique      also      adopted      in
Ref. \onlinecite{sherman2}.  We also  notice poor convergence  of 
self-consistent equations as $M_S$ becomes very small.

When    calculating    the    staggered   magnetization,    we    use
$C_1(k,\omega)=-D_I^{(11)}(k,\omega)/\pi$   for   both  positive   and
negative $\omega$. Numerically a Lorentzian of width $\eta$ is used to
describe the imaginary part of the bare Green's function $D_0^{(11)}$,
which actually should  be a delta function at  the non-negative energy
$\Omega_k$.  And, thus, to obtain the  normalization correctly,  we also
need  to consider  the tail  of the  Lorentzian even though that extends to
negative $\omega$.


We  note that  in  calculating the  imaginary  part of  the hole  
self-energy, the energy convolution is  utilized to get rid of the infinite
integration  limits.   However  in  that   derivation  the  Lorentzian
approximation   of   the   delta   function   overlooks   the   finite
Lorentzian-width coming from the finite $\eta$.


\begin{figure}[htbp]
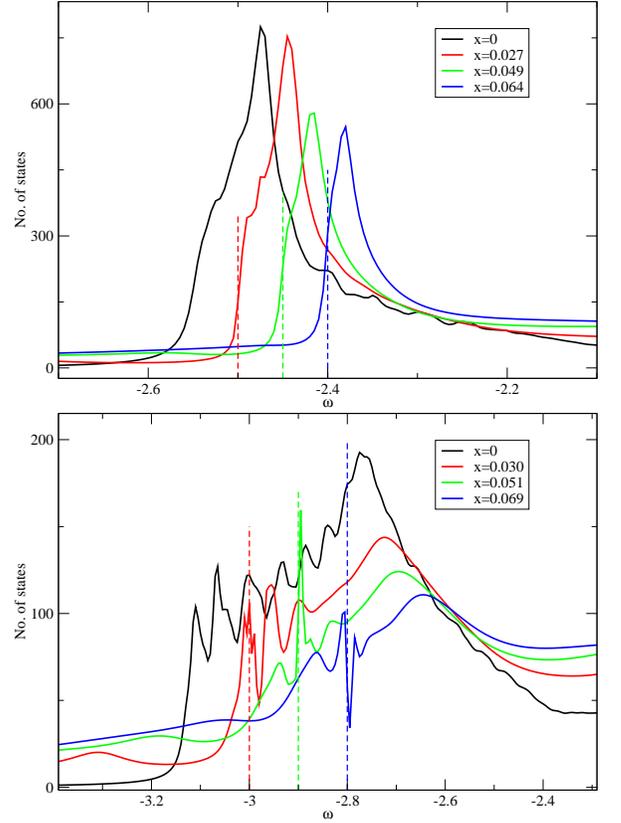

\begin{center}
\epsfig{file=Fig20a.eps,width=.9\linewidth,clip=}\\ \epsfig{file=Fig20b.eps,width=.9\linewidth,clip=}
\end{center}
\caption{Hole   density   of   states (number of states in each of our
$\delta \omega$)   in   the   $t-J$   (top)   and
  $t-t^\prime-t^{\prime\prime}-J$   (bottom)   models   for   $J=0.2t$
  (L=32). The positions of $\mu$ are also shown.}
\label{figd4}
\end{figure}
\begin{figure}[htp]
\vskip .15 in
\begin{center}
\includegraphics[width=.9\linewidth]{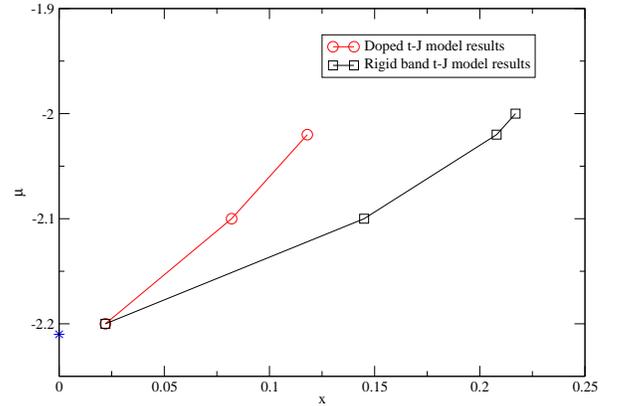}
\end{center}
\caption{Variation of $\mu$ with doping concentration $x$ for $J=0.4t$
  in a $32  \times 32$ lattice. The blue star on  the y-axis at energy
  -2.21$t$  shows  the quasi-particle  peak  position  of the  undoped
  antiferromagnet.}
\label{figd20}
\end{figure}

\begin{table*}
\begin{center}
\caption{Doping concentrations  at different $\mu$  for different sets
  of $\eta$}
\begin{tabular}  {|p{30pt}| p{40.00pt}| p{40.00pt}| p{40.00pt}|
    p{40.00pt}|     p{40.00pt}|     p{40.00pt}|p{40.00pt}|}     \hline
  $\eta\backslash\mu$ & -2.7 & -2.5 & -2.4 & -2.3 & -2.2 & -2.1 & -2.0
  \\ \hline 0.1 & 0.0028 & 0.0041 & 0.0053 & 0.0080 & 0.0269 & 0.142 &
  0.216 \\ \hline 0.01 & 0.0019 &  0.0027 & 0.0034 & 0.0053 & 0.0218 &
  0.145 & 0.217 \\ \hline
\end{tabular} 
\label{tabled}
\end{center}
\end{table*}

We find that  the density of states below  the quasi-particle peak, as
obtained from the numerical calculation, depends strongly on the value
of the broadening parameter, $\eta$ used. This is an error which comes
as a  result of the finite  $\eta$ used in  numerical calculations but
this $\eta$-dependence reduces significantly as higher values of Fermi
energy  (i.e., higher  doping concentrations)  are  considered.  Table
\ref{tabled}  illustrates the  result for  $J=0.4t$ on  a  $t-J$ model
rigid band. The numerical results coming from a system where the Fermi
level stays much  higher than the quasi-particle energy  has a smaller
error due  to finiteness of  the $\eta$-used. While studying  the hole
energy bands from the doped  system we have, thus, considered only the
cases where  the Fermi energy  is higher than the  quasi-particle peak
position (which is at $\sim-2.21t$ for $J=0.4t$).

 The  density of states (DOS)  at low energies  (close to Fermi
energy   of   the   undoped   system)  becomes   modified   with   doping
(Fig. \ref{figd4})  and the DOS  low energy peaks move  towards higher
values at  higher dopings  (also see Ref.  \onlinecite{sherman}). This
shifting of the  low energy states towards higher  energies causes the
chemical potential to  change rapidly with doping as  compared to that
calculated   on    a   single   hole   rigid   band    as   shown   in
Fig. \ref{figd20}.  In the  range of  doping in which  we are  able to
study the doped system within the realm of the LSW approximation we do
not see $\mu$  to cross the DOS  low energy peak for the  $t-J$ model. In
the $t-t^\prime -t^{\prime\prime}-J$ model, on the other hand, the low
energy peaks in the DOS are much  flatter and $\mu$ is seen to cross a
few of them even at low doping (Fig. \ref{figd4}).

As was discussed in the introduction our results differ
significantly from those of Ref.~\onlinecite{sherman}. For example,
we find that the staggered magnetization decreases very rapidly with 
doping becoming zero for $x \sim 0.08$ (at $J/t=0.2$).
In contrast, in Ref.~\onlinecite{sherman} results  were presented 
for much larger values of doping concentration $x$  ($\sim 0.25$),
where according to our calculation, this type of calculation, 
which requires presence of antiferromagnetic long-range order, is invalid.
The dependence of our magnon spectra with doping is also consistent with
the vanishing of the staggered magnetization at $x \sim 0.08$. Namely,
the renormalized magnon energy softens with doping,  
for    small     $k$    values, becoming practically zero for  
   $x\sim    0.08$    (for $J=0.2t$).  

Because of the above disagreement we subjected our code to several tests.
We calculated
  the  self-energies  for a  single iteration using our program
and the result agrees  with the results which we 
obtain using the  mathematical expressions obtained  
analytically. Second our results for the
  spectra  at very  small doping  matches  well with  the single  hole
  spectra (as  computed in  Ref. \onlinecite{liu}) at  energies higher
  than the quasi-particle peak of the undoped antiferromagnet. We also
  check the  dependence of  energy resolution ($\Delta\omega$)  on our
  numerical    integration   ($I_{\Delta\omega}$)   and    find   that
  $I-I_{\Delta\omega}\sim(\Delta\omega)^2$, as it should be.

\end{document}